\documentclass[fleqn,usenatbib]{mnras}

\usepackage{newtxtext,newtxmath}

\usepackage[T1]{fontenc}

\DeclareRobustCommand{\VAN}[3]{#2}
\let\VANthebibliography\thebibliography
\def\thebibliography{\DeclareRobustCommand{\VAN}[3]{##3}\VANthebibliography}

\usepackage{graphicx}
\usepackage{subcaption}
\usepackage{amsmath}	
\definecolor{orcidlogocol}{HTML}{A6CE39}
\newcommand{\orcid}[1]{\href{https://orcid.org/#1}{\includegraphics[width=8pt]{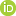}}}

\title[TOI-6884b: a transiting brown dwarf]%
{TOI-6884b: A low-mass brown dwarf transiting a slightly evolved star}
\author[Akanksha Khandelwal et al.]{
Akanksha Khandelwal$^{1,2}$\thanks{E-mail: akanksha@astro.unam.mx}\orcid{0000-0003-0335-6435},
Shubhendra Nath Das$^{2,3}$\orcid{0009-0006-9996-1814},
Rishikesh Sharma$^{2}$\orcid{0000-0001-8983-5300},
Abhijit Chakraborty$^{2}$\orcid{0000-0002-3815-8407},
\newauthor
Churchil Dwivedi$^{2}$\orcid{0000-0002-8804-650X},
Sanjay Baliwal$^{2}$\orcid{0000-0001-8998-3223},
Karen A.\ Collins$^{4}$\orcid{0000-0001-6588-9574},
David W.\ Latham$^{4}$\orcid{0000-0001-9911-7388},
Allyson Bieryla$^{4}$\orcid{0000-0001-6637-5401},
\newauthor
Cristilyn N.\ Watkins$^{4}$\orcid{0000-0001-8621-6731},
Felipe Murgas$^{5,6}$\orcid{0000-0001-9087-1245},
Norio Narita$^{5,7,8}$\orcid{0000-0001-8511-2981},
Enric Pall\'e$^{5,6}$\orcid{0000-0003-0987-1593},
Steve B.\ Howell$^{9}$\orcid{0000-0002-2532-2853},
\newauthor
Mark E.\ Everett$^{10}$\orcid{0000-0002-0885-7215},
Catherine A.\ Clark$^{11}$\orcid{0000-0002-2361-5812},
Polina A.\ Budnikova$^{12}$\orcid{0009-0002-0733-572X},
David Ciardi$^{11}$\orcid{0000-0002-5741-3047},
Nikitha Jithendran$^{2}$\orcid{0009-0000-4834-5612},
\newauthor
Akihiko Fukui$^{5,7}$\orcid{0000-0002-4909-5763},
Ashirbad Nayak$^{2}$\orcid{0009-0001-3782-4308},
Bob Massey$^{13}$\orcid{0000-0001-8879-7138},
Boris Safonov$^{12}$\orcid{0000-0003-1713-3208},
Florence Libotte$^{5,17}$\orcid{0009-0004-3455-5140},
\newauthor
Francis P.\ Wilkin$^{14}$\orcid{0000-0003-2127-8952},
Gregor Srdoc$^{15}$,
Howard M.\ Relles$^{4}$\orcid{0009-0009-5132-9520},
Ivan Bonilla-Mariana$^{5,6}$,
Izuru Fukuda$^{19}$\orcid{0000-0002-9436-2891},
\newauthor
Jason D.\ Eastman$^{4}$\orcid{0000-0003-3773-5142},
Jerome de Leon$^{7}$\orcid{0000-0002-6424-3410},
Jesus Higuera$^{10}$\orcid{0000-0002-3985-8528},
Kapil Bharadwaj$^{2}$\orcid{0000-0003-1373-4583},
Keith Horne$^{16}$\orcid{0000-0003-1728-0304},
Kendra Nguyen$^{18}$\orcid{0000-0002-5937-9655},
\newauthor
Kevikumar Lad$^{2}$\orcid{0009-0008-4890-9527},
Manuel Pichardo Marcano$^{1}$\orcid{0000-0003-4436-831X},
Micaela Magno$^{14}$\orcid{0009-0005-5848-9376},
Neelam JSSV Prasad$^{2}$\orcid{0000-0003-0670-5821},
\newauthor
Noriharu Watanabe$^{19}$\orcid{0000-0002-7522-8195},
Richard P.\ Schwarz$^{4}$\orcid{0000-0001-8227-1020},
Sam Quinn$^{4}$\orcid{0000-0002-8964-8377},
Santiago P\'aez$^{1}$\orcid{0000-0001-6381-7120},
Toshi Suganuma$^{19}$\orcid{0009-0007-2926-1924},
\newauthor
Y.\ G\'omez Maqueo Chew$^{1}$\orcid{0000-0002-7486-6726}
\\\\
$^{1}$ Universidad Nacional Aut\'onoma de M\'exico. Instituto de Astronom\'ia. A.P. 70-264, 04510. Ciudad de M\'exico, M\'exico.\\
$^{2}$ Astronomy \& Astrophysics Division, Physical Research Laboratory, Ahmedabad 380009, India\\
$^{3}$ Indian Institute of Techonology, 382355 Gandhinagar, India\\
$^{4}$ Center for Astrophysics | Harvard \& Smithsonian, 60 Garden Street, Cambridge, MA 02138, USA\\
$^{5}$ Instituto de Astrof\'isica de Canarias (IAC), 38205 La Laguna, Tenerife, Spain\\
$^{6}$ Departamento de Astrof\'isica, Universidad de La Laguna (ULL), E-38206 La Laguna, Tenerife, Spain\\
$^{7}$ Komaba Institute for Science, The University of Tokyo, 3-8-1 Komaba, Meguro, Tokyo 153-8902, Japan\\
$^{8}$ Astrobiology Center, 2-21-1 Osawa, Mitaka, Tokyo 181-8588, Japan\\
$^{9}$ NASA Ames Research Center, Moffett Field, CA 94035, USA\\
$^{10}$ NSF National Optical-Infrared Astronomy Research Laboratory, 950 N. Cherry Ave., Tucson, AZ 85719, USA\\
$^{11}$ NASA Exoplanet Science Institute, Caltech/IPAC, Pasadena, CA 91125, USA\\
$^{12}$ Lomonosov Moscow State University, Sternberg Astronomical Institute, Moscow, Russia\\
$^{13}$ Villa '39 Observatory, Landers, CA 92285, USA\\
$^{14}$ Department of Physics and Astronomy, Union College, 807 Union St., Schenectady, NY 12308, USA\\
$^{15}$ Kotizarovci Observatory, Sarsoni 90, 51216 Viskovo, Croatia\\
$^{16}$ SUPA Physics and Astronomy, University of St. Andrews, Fife, KY16 9SS, Scotland, UK\\
$^{17}$ Sabadell Astronomical Society, Sabadell, Spain\\
$^{18}$ Department of Astronomy, Yale University, 219 Prospect Street, New Haven, CT 06511, USA\\
$^{19}$ Department of Multi-Disciplinary Sciences, Graduate School of Arts and Sciences, The University of Tokyo, 3-8-1 Komaba, Meguro, Tokyo 153-8902, Japan
}
\date{Accepted XXX. Received YYY; in original form ZZZ}

\pubyear{\the\year{}}
\providecommand{\bjdtdb}{\ensuremath{\rm {BJD_{TDB}}}}

\providecommand{\msun}{\ensuremath{M_\odot}}
\providecommand{\rsun}{\ensuremath{R_\odot}}
\providecommand{\lsun}{\ensuremath{L_\odot}}
\providecommand{\msun}{\ensuremath{M}}
\providecommand{\rsun}{\ensuremath{R}}
\providecommand{\lsun}{\ensuremath{L}}
\providecommand{\mj}{\ensuremath{\,M_{\rm J}}}
\providecommand{\rj}{\ensuremath{\,R_{\rm J}}}

\providecommand{\fave}{\langle F \rangle} 
\providecommand{\fluxcgs}{10$^9$ erg s$^{-1}$ cm$^{-2}$}

\providecommand{\arcsec}{$^{\prime \prime}$}
\providecommand{\arcmin}{$^{\prime}$}

\begin{document}
\label{firstpage}
\pagerange{\pageref{firstpage}--\pageref{lastpage}}
\maketitle

\begin{abstract}
We report the discovery of a low-mass transiting brown dwarf orbiting TOI-6884 (TIC~156514476, $T_{\rm mag}=11.4$) from NASA’s \textit{Transiting Exoplanet Survey Satellite} (\textit{TESS}) mission. The \textit{TESS} light curves initially suggested an orbital period of $\sim$14.42~days; however, our high-precision ground-based radial velocity measurements and multi-epoch time-series photometry reveal this to be a harmonic alias. We determine the true orbital period to be $4.808264^{+0.000015}_{-0.000014}$~days and confirm the substellar nature of the companion. TOI-6884b has a mass of $26.32^{+0.98}_{-0.93}$\mj, a radius of $0.927^{+0.51}_{-0.52}$ \rj, and resides on a nearly circular orbit ($e=0.067^{+0.010}_{-0.012}$). Its host star is a late F-type slightly evolved star with $M_\star = 1.410^{+0.075}_{-0.069}$ \msun, $R_\star = 1.840^{+0.072}_{-0.073}$ \rsun, $\log{g} = 4.057^{+0.045}_{-0.039}$, $[{\rm Fe/H}] = 0.094^{+0.073}_{-0.068}$~dex, and $T_{\rm eff}=6330^{+180}_{-160}$ K. TOI-6884b is a key addition to the small population of well-characterized transiting brown dwarfs orbiting host stars that have left the main sequence. The detection of such systems will contribute to our understanding of dynamical histories and structural evolution of short-period substellar companions around evolved stars.
\end{abstract}




\section{Introduction}
Brown dwarfs (BDs) are substellar objects that bridge the gap between giant planets and the lowest-mass hydrogen-burning stars. They are commonly defined as occupying a mass range between the deuterium-burning limit at approximately 11--16\mj~and the onset of sustained hydrogen fusion near 75--80\mj~\citep{Spiegel2011,Baraffe2002,Chabrier2023}. The spread in these estimates can be explained by variations in metallicity, chemical composition, and formation conditions, reflecting the complex physics that governs their interiors. {For example, at high metallicity (+0.5 dex), the hydrogen-burning minimum mass (HBMM) decreases to $\sim$66\mj\ for hybrid atmosphere models and $\sim$70\mj\ for cloud-free atmospheres \citep{Morley2024}.}

Despite significant observational and theoretical efforts, the formation theories of BDs remain uncertain. It is suggested that they may form like stars via gravitational fragmentation of molecular clouds \citep{Padoan2004,Hennebelle2008,Kratter2016} or within protoplanetary discs via core accretion or gravitational instability, analogous to giant planets \citep{Alibert2005,Mordasini2009}. This raises the possibility that BDs do not form a homogeneous population but instead exhibit a diversity of origins. Observational demographics support this hypothesis and indicate that formation pathways can influence the orbital properties of BD companions. \citet{Ma2014} proposed a bifurcation in the BD population near $\sim$42.5\mj~based on orbital eccentricity distributions. Companions below this mass tend to show orbital architectures similar to giant planets \citep{Rodriguez2023}, whereas more massive BDs resemble stellar binaries in both eccentricity and period distributions \citep{Halbwachs2003,Kiefer2021} (although see \citealt{Schlaufman2018} for an alternative proposed boundary). Subsequent studies with larger samples have generally supported this trend \citep{Grieves2017,Grieves2021,Kiefer2021,Page2024}. In particular, the eccentricity distribution of short-period systems, where tidal interactions can circularize orbits, preserves crucial clues to the formation and migration history of these objects \citep{Hut1981,Adams2006,Jackson+2009,Beatty2018}.

Given this, well-characterized BD companions are therefore essential benchmarks for understanding substellar physics. Transiting BDs are especially valuable because they enable direct, model-independent measurements of both mass and radius, yielding stringent tests of substellar structure and evolution \citep{Burrows2011,Baraffe2015,Chabrier2000,Morley2024,Mukherjee2025}. Measurements of mass and radius probe the mass--radius ($M-R$) relation, which is sensitive to age, atmospheric composition, equation of state, and thermal evolution. Comparisons between empirical mass--radius measurements and theoretical predictions have revealed phenomena such as radius inflation, metallicity effects, and cooling inefficiencies in some systems \citep{Phillips2020,Page2024}. As BDs lack sustained nuclear fusion, they cool and contract over time, making their observed properties sensitive probes of substellar evolution. Each new system with precisely determined parameters contributes significantly to constraining substellar models and helps in disentangling the roles of formation environment and evolution.

An additional layer of complexity arises when the host star evolves off the main sequence {($\log{g}>4.1$ for main-sequence stars; \citealt{2018AJ....156..102S}).} The evolution of a BD companion in this regime becomes particularly intriguing, yet BDs orbiting evolved stars remain poorly characterized due to observational challenges introduced by the large radii and strong radial-velocity jitter of their hosts \citep{Yu2018,Tayar2019}. Moreover, companions of evolved stars are expected to follow different evolutionary pathways compared to those orbiting main-sequence hosts, owing to stronger dynamical interactions \citep{Veras2016}. For example, as the host star becomes more luminous, close-in companions receive increased irradiation, providing opportunities to study processes such as planetary inflation and re-inflation \citep[e.g.,][]{Grunblatt2016,Saunders2022}. At the same time, increased tidal forces can also cause faster orbital decay or even engulfment of the companion \citep{Hut1981,Jackson+2008} which may induce the different eccentricity distribution seen in surviving hot Jupiters \citep{Villaver2014,Grunblatt2018}. Compared to planetary companions, BDs around evolved stars remain especially unexplored because only a small number of systems {($\sim$11)} are known. Consequently, it is unclear whether BDs undergo orbital or structural changes analogous to those inferred for planets, highlights the need to expand the sample of well-characterized transiting BDs around evolved hosts.

In this work, we present the discovery and characterization of TOI-6884b, a transiting BD with a mass of $26.32^{+0.98}_{-0.93}$\mj, a radius of $0.927^{+0.51}_{-0.52}$\rj, and an age of $2.61^{+0.78}_{-0.75}$~$Gyr$, orbiting an F-type slightly evolved star with $T_{\rm eff}=6330^{+180}_{-160}$~K. The paper is organized as follows. In Section~\ref{sec:obs}, we describe the observations. Section~\ref{sec:analysis} presents our analysis and derived system parameters. In Section~\ref{sec:discussion}, we place TOI-6884b in the context of the known transiting BD population, with emphasis on the mass--radius relation and orbital eccentricity. Finally, we summarize our findings in Section~\ref{sec:summary}.
\section{Observations}\label{sec:obs}
\subsection{TESS photometry}
\begin{figure}
    \centering
    \begin{subfigure}{\linewidth}
        \centering
        \includegraphics[width=\linewidth]{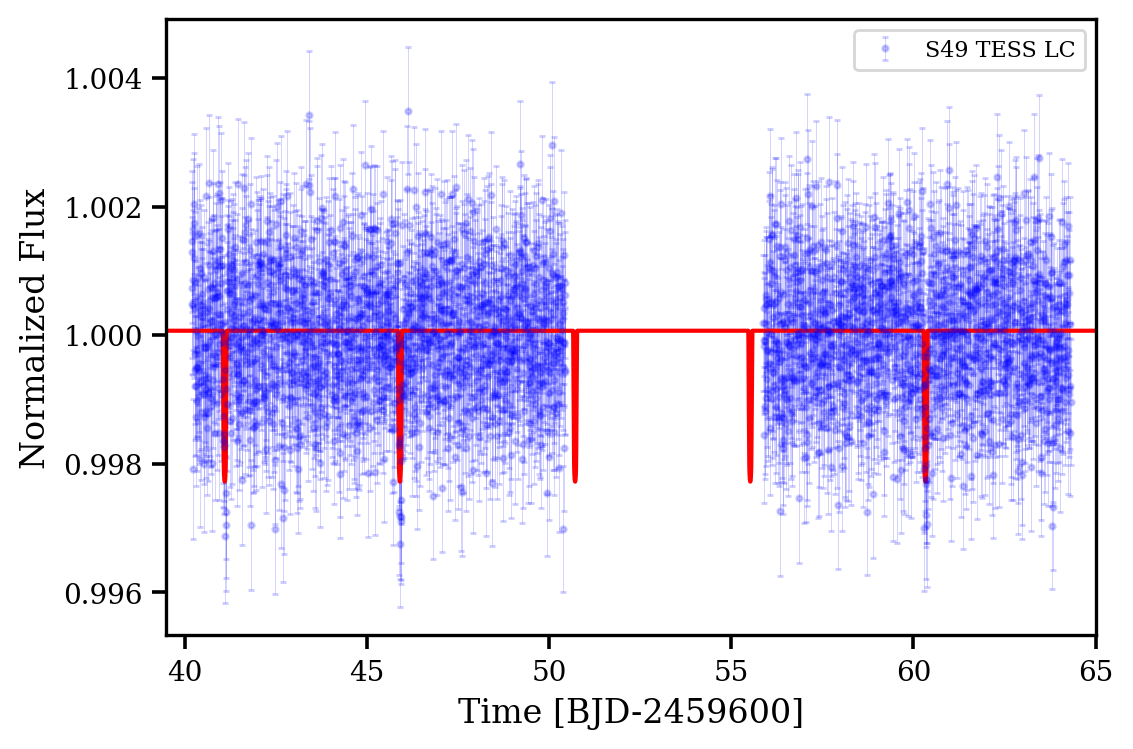}
    \end{subfigure}

    \begin{subfigure}{\linewidth}
        \centering
        \includegraphics[width=\linewidth]{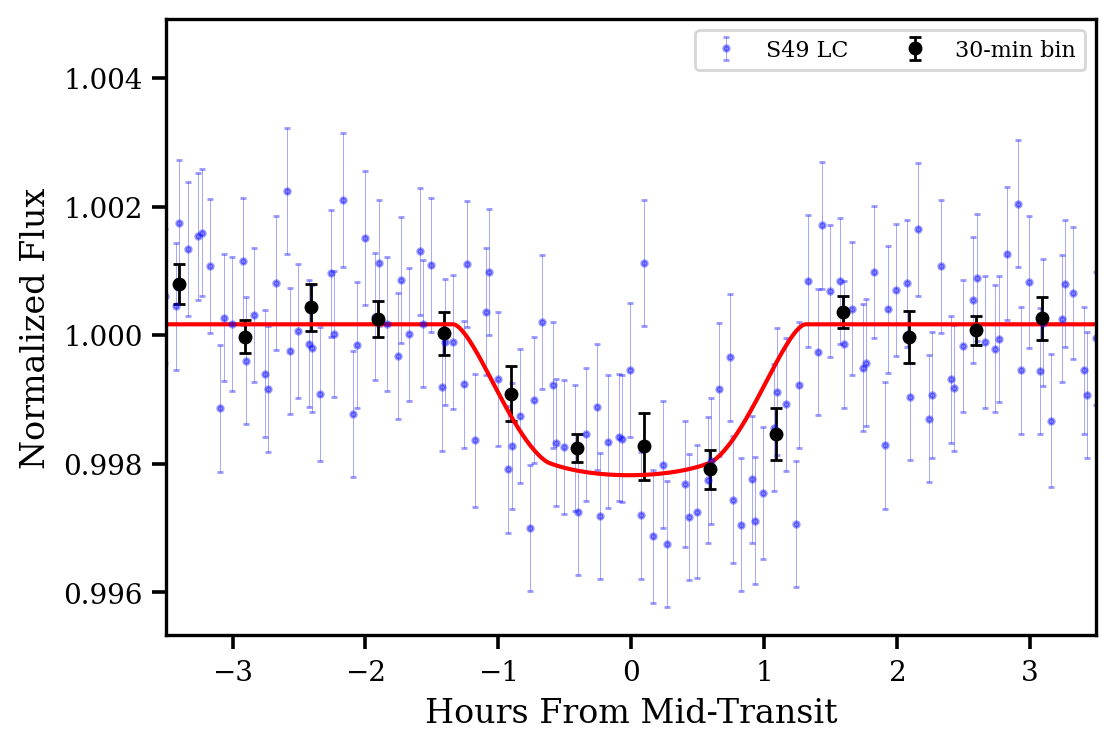}
    \end{subfigure}

    \caption{TESS light curve (LC) of TOI-6884 after detrending. The upper panel shows the full time-series LC, and the lower panel displays the phase-folded transit LC. Both panels include the best-fit transit model (red) and 30-minute binned data.}
    \label{fig:toi6884_transit}
\end{figure}

TOI-6884 (TIC~156514476, $T_{\rm mag}=11.4$) was observed by NASA’s \textit{Transiting Exoplanet Survey Satellite} (\textit{TESS}) in Sector 49 (UT 2022 February 26–March 25) with a 10-minute cadence on Camera 1 CCD 1. The star was initially flagged as a Community TESS Object of Interest \citep[CTOI;][]{Guerrero2021} by \citet{Nguyen2022}, based on transit signals identified in the light curves extracted from the TESS Full Frame Images (FFIs) of Sector 49 using the TESS Science Processing Operations Center (SPOC) pipeline \citep{spoc}, and was subsequently alerted as a TESS Object of Interest (TOI) on 2024 February 1. We used the Pre-search Data Conditioning Simple Aperture Photometry \citep[PDCSAP:][]{smith_2012,Stumpe_2014} light curves produced by the SPOC pipeline, which are publicly available via the Mikulski Archive for Space Telescopes (MAST)\footnote{\url{https://mast.stsci.edu/portal/Mashup/Clients/Mast/Portal.html}}. These light curves were normalized and detrended using a Python package \texttt{citlalicue} \citep{citlalicue}, which applies Gaussian Process regression to mitigate systematic effects and stellar variability. The resulting post-processed light curves are shown in Fig.~\ref{fig:toi6884_transit}. The \textit{TESS} pipeline initially reported an orbital period of $\sim$14.42~days. Upon careful inspection, we found that the first transit was present but not identified by the pipeline, producing a harmonic alias. Consequently, the true orbital period is approximately one-third of the initially reported value, as evident in the data (Fig.~\ref{fig:toi6884_transit}). A total of three transits are detectable in the data, while two additional events were missed due to spacecraft reorientation during data downlink.

\begin{figure}
    \centering
    \includegraphics[width=\columnwidth]{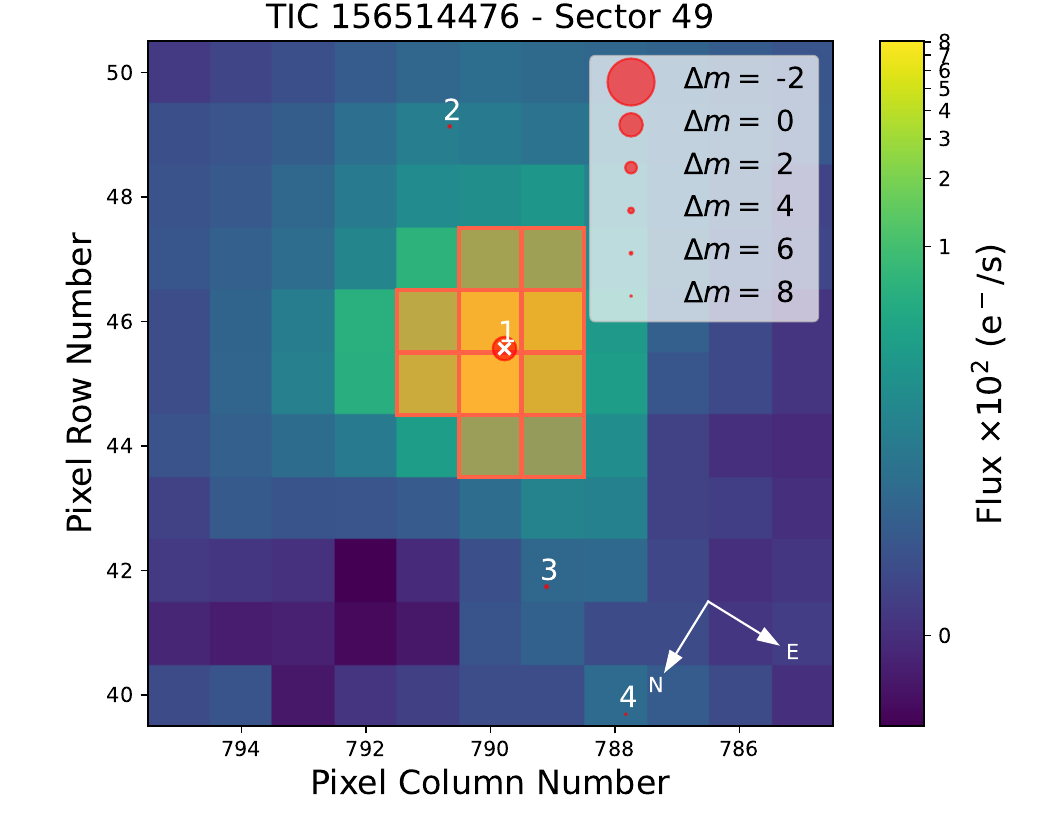}
    \caption{TESS target pixel file of TOI-6884 from Sector 49, generated with \texttt{tpfplotter} \citep{tpfplot}. 
    The target star is marked with a white cross and labeled as 1. Nearby Gaia DR3 sources within a magnitude contrast of $\Delta m \leq 8$ are shown as red circles, with symbol sizes scaled by relative brightness. 
    The SPOC photometric aperture is indicated by the shaded red pixels.}
    \label{fig:toi6884_tpfplot}
\end{figure}

Given the relatively large TESS pixel scale of 21\arcsec per pixel, contamination from nearby stars is possible. To assess this, we used the \texttt{tpfplotter} package \citep{tpfplot} and the \texttt{TESS-cont}\footnote{Available at \url{https://github.com/castro-gzlz/TESS-cont}.} algorithm \citep{CastroGonzalez2024b} to examine sources within the SPOC aperture. Nearby stars with magnitude contrasts up to 8 could, in principle, mimic the shallower transit of TOI-6884.02 \citep{LilloBox2014}. Fig.~\ref{fig:toi6884_tpfplot} shows the TESS target pixel file and the SPOC pipeline aperture for Sector 49. Accounting for the relative positions of nearby sources and the TESS pixel response functions (PRFs), \texttt{TESS-cont} estimates that 99.98\% of the flux in the aperture originates from TOI-6884. Therefore, the observed transit signal is intrinsic to the target star, and contamination from nearby sources can be confidently ruled out. Furthermore, the Gaia renormalized unit weight error (RUWE) for TOI-6884 is 1.22, consistent with a single-star astrometric solution \citep{gaia2023}.

To confirm the substellar nature of the companion and refine its orbital parameters, we conducted various ground-based follow-up observations, including high-precision radial velocity (RV) measurements and multi-epoch time-series photometry. The combination of \textit{TESS} photometry and ground-based observations allowed us to determine the accurate orbital period, measure the companion’s mass and radius, and rule out potential false positive scenarios.

\subsection{Ground-based Photometry\label{subsec:ground}}
\begin{figure}
    \centering
    \includegraphics[width=\columnwidth]{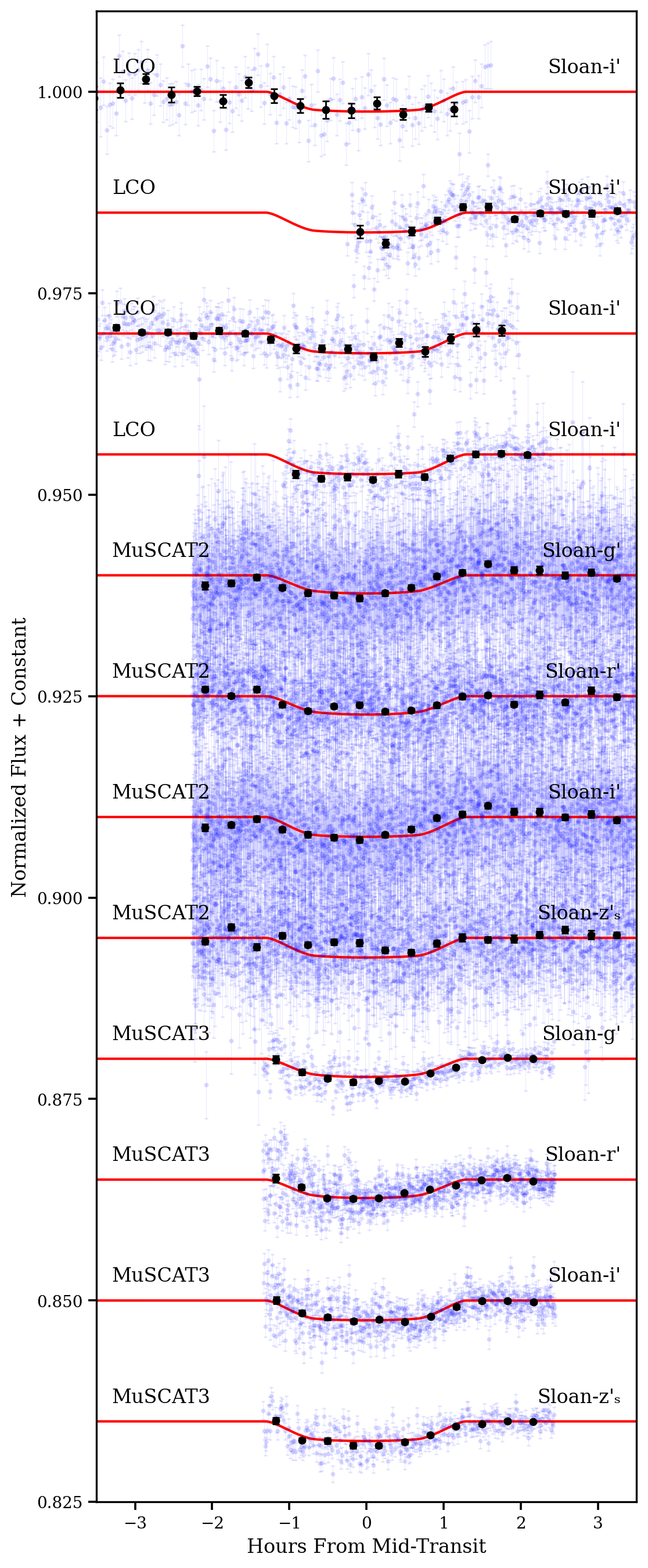}
    \caption{Transit light curves of TOI-6884 from our follow-up observations with MuSCAT-2 and LCOGT 1\,m, 0.35\,m, and 2\,m MuSCAT-3. All the dataset are 20-minutes binned. The solid red line represents the best-fit transit model generated with the \texttt{batman} package \citep{batman}, using the parameters from the most probable EXOFASTv2 solution (the high-mass fit, Pr $\sim77\%$). Details of the observational parameters are listed in Table \ref{ground_based_transit_table}.\label{fig:ground_based_transit}}
\end{figure}

Ground-based time-series follow-up photometry of the field around TOI-6884 was obtained as part of the TESS Follow-up Observing Program \citep[TFOP;][]{Collins2017}\footnote{https://tess.mit.edu/followup} to independently validate and refine the ephemerides of the 4.808-day orbital period. Transit observations were scheduled using the {\tt TESS Transit Finder}, a customized implementation of the {\tt Tapir} software package \citep{Jensen2013}. All light curve data are available on the {\tt EXOFOP-TESS} website\footnote{\href{https://exofop.ipac.caltech.edu/tess/target.php?id=156514476}https://exofop.ipac.caltech.edu/tess/target.php?id=156514476} and are summerized in Table \ref{ground_based_transit_table} and included in the global modelling described in section \ref{sec:global_modeling} and also see Fig. \ref{fig:ground_based_transit}.
\subsubsection{LCOGT}
We observed one full and two partial transit windows of TOI-6884.01 on UT 2024 April 5, 2024 April 24, and 2024 April 29 in Sloan $i'$ from the Las Cumbres Observatory Global Telescope \citep[LCOGT]{Brown2013} 1\,m network node at McDonald Observatory near Fort Davis, Texas, United States (McD). The 1\,m telescopes are equipped with a $4096\times4096$ SINISTRO camera having an image scale of $0\farcs389$ per pixel, resulting in a $26\arcmin\times26\arcmin$ field of view. We observed another partial transit window on UT 2024 February 12 from the LCOGT 0.35\,m network node at Teide Observatory on the island of Tenerife (TEID). The 0.35\,m Planewave Delta Rho 350 telescope is equipped with a $9576\times6388$ QHY600 CMOS camera having an image scale of $0\farcs73$ per pixel, resulting in a $114\arcmin\times72\arcmin$ field of view. However, the image data were collected using the $30\arcmin\times30\arcmin$ central field of view sub-frame mode. Finally, we observed a full transit with limited pre-ingress baseline coverage on UT 2024 December 25 from the LCOGT 2\,m Faulkes Telescope North at Haleakala Observatory on Maui, Hawai'i. The 2\,m LCOGT telescope is equipped with the MuSCAT3 multi-band imager \citep{Narita:2020}. All images were calibrated by the standard LCOGT {\tt BANZAI} pipeline \citep{McCully2018}, and differential photometric data were extracted using {\tt AstroImageJ} \citep{Collins2017}. We used circular photometric apertures with radii $4\farcs3$ to $7\farcs3$ that excluded all of the flux from the nearest known neighbor in the Gaia DR3 catalog (Gaia DR3 3961540708410542592) that is $71\farcs1$ southeast of TOI-6884. 

\subsubsection{MuSCAT2}
A full transit of TOI-6884.01 was observed on UT 2024 May 3 with MuSCAT2 \citep{Narita2019}, a four-channel imager mounted on the 1.5\,m Telescopio Carlos S\'{a}nchez (TCS) at the Teide Observatory, Spain. MuSCAT2 enables simultaneous imaging in the $g'$, $r'$, $i'$, and $z_s$ bands using four $1024 \times 1024$ CCDs, each covering a field of view of $7.4$\arcmin $\times$ $7.4$\arcmin.

The observations were carried out with the telescope in nominal focus. Exposure times of 8\,s in $g'$ and 10\,s in $r'$, $i'$, and $z_s$ were adopted. The data were reduced using the MuSCAT2 pipeline \citep{Parviainen2019}, which performs dark subtraction, flat-field correction, aperture photometry, and a simultaneous fit of the transit model and instrumental systematics, while selecting the best aperture for the time series. The transit signal was clearly detected in the photometry of the target star.

\begin{table*}
    \caption{A summary of the ground-based transit follow-up observations}\label{ground_based_transit_table}
    \centering
    \begin{tabular}{llllll}
        \hline\hline
        Target & Telescope & Date & Filter & Exposure Time (s) & Photometric Aperture ($''$) \\
        \hline
        TOI-6884.01 & LCO/Teid     & Feb 11 2024 & Sloan-$i'$          & 155 & 8.1  \\
        TOI-6884.01 & LCO/McD      & Apr 5 2024  & Sloan-$i'$          & 52  & 5.8  \\
        TOI-6884.01 & LCO/McD      & Apr 24 2024 & Sloan-$i'$          & 54  & 4.3  \\
        TOI-6884.01 & LCO/McD      & Apr 29 2024 & Sloan-$i'$          & 52  & 3.9  \\
        TOI-6884.01 & TCS/MuSCAT-2 & May 3 2024 & Sloan-$g'r'u'z_s'$ & 11  & 10.9 \\
        TOI-6884.01 & LCO-HAL/MuSCAT-3 & Dec 25 2024 & Sloan-$g'r'u'z_s'$ & 53  & 7.2 \\
        \hline
    \end{tabular}
\end{table*}


\subsection{PARAS-2 radial velocities}

The RV observations were taken using the PARAS-2 \citep{Chakraborty2018, Chakraborty2024} high-resolution spectrograph attached to PRL 2.5\,m telescope situated at the Mount Abu Observatory, Rajasthan, India. PARAS-2 is a high-resolution ($R$ $\approx$ 110,000) echelle-based spectrograph maintained in a highly temperature- and pressure-stable environment and works in the wavelength range of 380-690 nm. This spectrograph uses simultaneous observations from Uranium-Argon hollow cathode lamps for precise measurements of instrumental drifts. This instrument has shown an on-sky stability of 2.65 m\,s$^{-1}$ over more than a month on a RV standard star \citep{Baliwal2024}. We have computed the RV values by performing cross-correlation of the observed spectrum with a template spectrum \cite{Baranne1996}. One can refer to \cite{Baliwal2024} for a detailed discussion of the data reduction and the analysis of the science observations taken with PARAS-2. We have taken a total of 20 spectroscopic observations with 3600 seconds of exposure time, spanning over 68 days from 2024 February 15 to 2024 April 22. In those 26 spectroscopic observations, we have obtained per pixel signal to noise ratios ranging from 9 to 30 at 550 nm with a median value of 14. The photon noises in the RVs range from 48.22 m\,s$^{-1}$ to 167.58 m\,s$^{-1}$, with a median value of 83.15 m\,s$^{-1}$. The calculation of the photon noise is based on the methods mentioned in \cite{Chaturvedi2016}. All the details of the RV observations for TOI-6884 are listed in Table \ref{tab:RV_measurements}.

\begin{table*}
\begin{center}
\caption{Radial velocity measurements of TOI-6884 from PARAS-2 and TRES. Column 1 lists the barycentric Julian dates (BJD$_\mathrm{TDB}$) of the observations, followed by the relative RVs and their associated uncertainties (Columns 2 and 3). Columns 4 and 5 give the bisector velocity spans and their corresponding uncertainties. The exposure times are provided in Column 6, and the final column identifies the instrument used for each observation.}
\label{tab:RV_measurements}
\begin{tabular}{lrrrrrl}
\hline\hline
BJD$_\mathrm{TDB}$ & \multicolumn{1}{c}{RV} & \multicolumn{1}{c}{$\sigma_{\mathrm{RV}}$} & 
\multicolumn{1}{c}{BIS} & \multicolumn{1}{c}{$\sigma_{\mathrm{BIS}}$} &
\multicolumn{1}{c}{Exp. time} & Instrument \\

(days) & (m\,s$^{-1}$) & (m\,s$^{-1}$) & (m\,s$^{-1}$) & (m\,s$^{-1}$) & (s) & \\ 
\hline
2460356.475724 & 472.99  & 42.75  & $-938.47$ & 109.53  & 3600 & PARAS-2 \\
2460368.474969 & $-4475.52$ & 68.77  & $-24.67$  & 143.20 & 3600 & PARAS-2 \\
2460376.479988 & $-1240.04$ & 103.68 & $-1218.89$ & 422.15 & 3600 & PARAS-2 \\
2460383.418293 & $-3739.71$ & 77.87  & $-304.96$ & 278.01 & 3600 & PARAS-2 \\
2460384.413929 & $-1215.37$ & 81.47  & 756.28   & 243.66 & 3600 & PARAS-2 \\
2460386.405985 & $-2548.83$ & 149.11 & $-442.40$ & 288.97 & 3600 & PARAS-2 \\
2460387.426305 & $-4238.95$ & 77.03  & $-577.17$ & 415.28 & 3600 & PARAS-2 \\
2460388.305673 & $-3871.34$ & 110.86 & $-112.64$ & 244.50 & 3600 & PARAS-2 \\
2460392.369709 & $-4052.31$ & 118.13 & $-1232.28$ & 552.41 & 3600 & PARAS-2 \\
2460398.354432 & $-2420.79$ & 167.58 & $-2000.00$ & 519.55 & 3600 & PARAS-2 \\
2460401.213912 & $-3730.36$ & 90.40  & $-1495.61$ & 534.65 & 3600 & PARAS-2 \\
2460407.305667 & $-4523.26$ & 99.05  & 311.32   & 842.42 & 3600 & PARAS-2 \\
2460408.247688 & $-1933.01$ & 94.35  & $-2781.35$ & 577.45 & 3600 & PARAS-2 \\
2460408.320533 & $-2545.80$ & 61.43  & $-1842.09$ & 178.89 & 3600 & PARAS-2 \\
2460409.343806 & 591.15  & 65.18  & $-1783.32$ & 217.96 & 3600 & PARAS-2 \\
2460419.260935 & $-1179.37$ & 48.22  & $-414.71$ & 254.16 & 3600 & PARAS-2 \\
2460419.311072 & $-572.84$ & 63.52  & 817.49   & 195.06 & 3600 & PARAS-2 \\
2460420.202738 & $-2381.70$ & 102.58 & $-205.40$ & 1431.03 & 3600 & PARAS-2 \\
2460420.293521 & $-2740.45$ & 66.52  & 136.94   & 187.33 & 3600 & PARAS-2 \\
2460423.319249 & $-493.88$ & 84.82  & 1398.94  & 530.77 & 3600 & PARAS-2 \\2460355.881282 & $1784.20$ & 47.80 & \ldots & \ldots & 720 & TRES \\
2460356.934107 & $1818.80$ & 36.80 & \ldots & \ldots & 960 & TRES \\
2460362.822458 & $-1659.80$ & 41.60 & \ldots & \ldots & 1200 & TRES \\
2460374.862271 & $1200.40$ & 24.60 & \ldots & \ldots & 1050 & TRES \\
2460386.849183 & $-1713.50$ & 40.10 & \ldots & \ldots & 1400 & TRES \\
2460387.878980 & $-2319.90$ & 32.30 & \ldots & \ldots & 800 & TRES \\
2460388.817584 & $-41.30$ & 33.50 & \ldots & \ldots & 900 & TRES \\
2460390.865991 & $1149.50$ & 34.80 & \ldots & \ldots & 900 & TRES \\
2460399.774034 & $2327.40$ & 33.60 & \ldots & \ldots & 1700 & TRES \\
2460403.824561 & $1476.80$ & 52.00 & \ldots & \ldots & 900 & TRES \\
2460405.916145 & $-1237.90$ & 40.80 & \ldots & \ldots & 1250 & TRES \\
2460408.813021 & $1867.70$ & 42.40 & \ldots & \ldots & 1500 & TRES \\

\hline
\end{tabular}
\end{center}

\end{table*}

\begin{table}[ht!]
\centering
\caption{Astrometry, photometry, and kinematics properties of TOI-6884.}
\begin{tabular}{lll}
\hline
\noalign{\smallskip}
 Parameter&Value& Ref.\\
\noalign{\smallskip}
\hline
\noalign{\smallskip}
Identifiers:\\
TIC&156514476&(1)\\
2MASS&J12520212+2640262&(3)\\
GaiaDR3&3961540914568972928& (4)\\
\noalign{\smallskip}
Astrometry:\\
$\alpha_{J2000}$ & 12:52:02.11 & (4)\\
$\delta_{J2000}$ & 26:40:26.22 & (4)\\
$\mu_{\alpha}$ (mas yr$^{-1}$) & 10.544 $\pm$ 0.022 & (4)\\
$\mu_{\delta}$ (mas yr$^{-1}$) & -21.279 $\pm$ 0.017 & (4)\\
$\varpi^*$ (mas) &$4.727\pm0.021$ &(4)\\
$d$ (pc) &$211.53^{+0.94}_{-0.93}$ &(4)\\
\noalign{\smallskip}
Photometry$^*$:\\
$B_{T}$ & 11.024  $\pm$ 0.046 & (2)\\
$V_{T}$ & 10.315  $\pm$ 0.034 & (2)\\
$T$   & 9.6089 $\pm$ 0.0062 & (1)\\
$G$   & 10.045  $\pm$ 0.020 & (4)\\
$G_{BP}$   & 10.370 $\pm$ 0.020 & (4)\\
$G_{RP}$   & 9.558 $\pm$ 0.020 & (4)\\
$J$   & 9.022 $\pm$  0.029 & (3)\\
$H$   & 8.736 $\pm$  0.031 & (3)\\
$K_{S}$ & 8.684 $\pm$ 0.020 & (3)\\
$W1$  &  8.648 $\pm$ 0.030 & (5)\\
$W2$  &  8.691 $\pm$ 0.030 & (5)\\
$W3$  &  8.650 $\pm$ 0.030 & (5)\\
$W4$  &  8.658 $\pm$ 0.288 & (5)\\
\noalign{\smallskip}
Kinematics:\\
$U,V,W$ ($km \ s^{-1}$) &  -31.203, -4.316, -13.763 & (6)\\
$U_{LSR},V_{LSR},W_{LSR}$ ($km \ s^{-1}$) &  -20.103,  7.924, -6.513 & (6)\\
\noalign{\smallskip}
\hline
\noalign{\smallskip}
\multicolumn{3}{l}{\footnotesize{$^*$A systematic error floor has been applied to the uncertainties.}}\\
\multicolumn{3}{l}{\footnotesize{References: (1) \cite{2018AJ....156..102S}, (2) \cite{tycho},}}\\
\multicolumn{3}{l}{\footnotesize{ (3) \cite{JHK}, (4) \cite{gaia2023},}}\\
\multicolumn{3}{l}{\footnotesize{ (5) \cite{ALLWISE}, (6) This work}}\\
\end{tabular}
\label{tab:star_table}
\end{table}

\subsection{TRES radial velocities}
The RV observations were obtained between February and April 2024 using the Tillinghast Reflector Echelle Spectrograph \citep[TRES;][]{gaborthesis}, mounted on the 1.5 m Tillinghast Reflector telescope at the Fred Lawrence Whipple Observatory (FLWO) atop Mount Hopkins, Arizona. TRES is an optical, fiber-fed echelle spectrograph with a wavelength coverage of 390–910 nm and a resolving power of $R \sim 44{,}000$. The spectra were obtained in sequences of three exposures bracketed by ThAr calibration spectra, and the median of each sequence was combined to remove cosmic rays. The average exposure time was $\sim$1100 s, resulting in an average signal-to-noise ratio per resolution element of 36.1. The spectra were extracted following the procedure described in \citet{Buchhave2010}. A multi-order analysis was then performed by cross-correlating each spectrum, order by order, against a co-added template spectrum to determine the RVs. The RVs derived from the TRES spectra, along with their associated uncertainties, are listed in Table~\ref{tab:RV_measurements}.

\subsection{High-angular-resolution imaging}\label{subsec:high_imaging}
If a host star has a spatially close companion, that companion (bound or line of sight, e.g., an EB) can create a false-positive transit signal. ``Third-light” flux from the close companion star can lead to an underestimated planetary radius if not accounted for in the transit model \citep{Ciardi2015}, cause non-detections of small planets residing with the same exoplanetary system \citep{Lester2021}, and cause the derivation of incorrect exoplanet and star properties \citep{Furlan2017,Furlan2020}. Additionally, the discovery of close, bound companion stars, which exist in nearly one-half of FGK type stars \citep{Matson2018}, provides crucial information toward our understanding of exoplanetary formation, dynamics and evolution \citep{Howell2021}. Thus, to search for close-in bound companions unresolved in TESS observations, we performed ground-based high-resolution follow-up observations of TOI-6884.

\subsubsection{Gemini observations}
\begin{figure}
    \centering
    \includegraphics[width=\columnwidth]{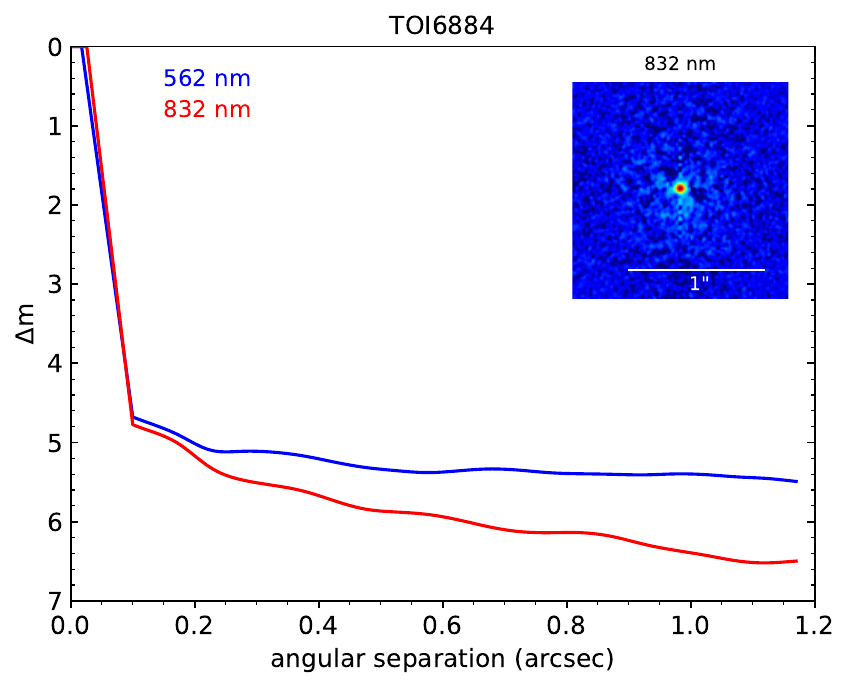}
    \caption{Speckle imaging observations of TOI-6884 obtained with ‘Alopeke at the Gemini North telescope. The magnitude contrast curves represent fits to the 5$\sigma$ detection limits from the diffraction limit out to the edge of the field of view. The inset panel shows the reconstructed image at 832\,nm.}
    \label{fig:toi6884_gemini}
\end{figure}

TOI-6884 was observed on UT 2024 March 22 using the ‘Alopeke speckle instrument on the Gemini North 8\,m telescope \citep{Scott2021}. ‘Alopeke provides simultaneous speckle imaging in two bands (562 nm and 832 nm) with output data products including a reconstructed image with robust contrast limits on companion detections. Nine sets of 1000 X 0.06 seconds exposures were collected and subjected to Fourier analysis in our standard reduction pipeline \citep[see][]{Howell2011}. Fig. \ref{fig:toi6884_gemini} shows our final contrast curves and the 832 nm reconstructed speckle image. We find that TOI-6884 is a single star with no companion star brighter than the 4.5 to 6.5 5$\sigma$ magnitude contrast below that of the target star from the diffraction limit (20 mas) out to 1.2\arcsec. At the distance of TOI-6884 (d=562 pc) these angular limits correspond to spatial limits of 11 to 674 au.

\subsubsection{Palomar observations}
\begin{figure}
    \centering
    \includegraphics[width=\columnwidth]{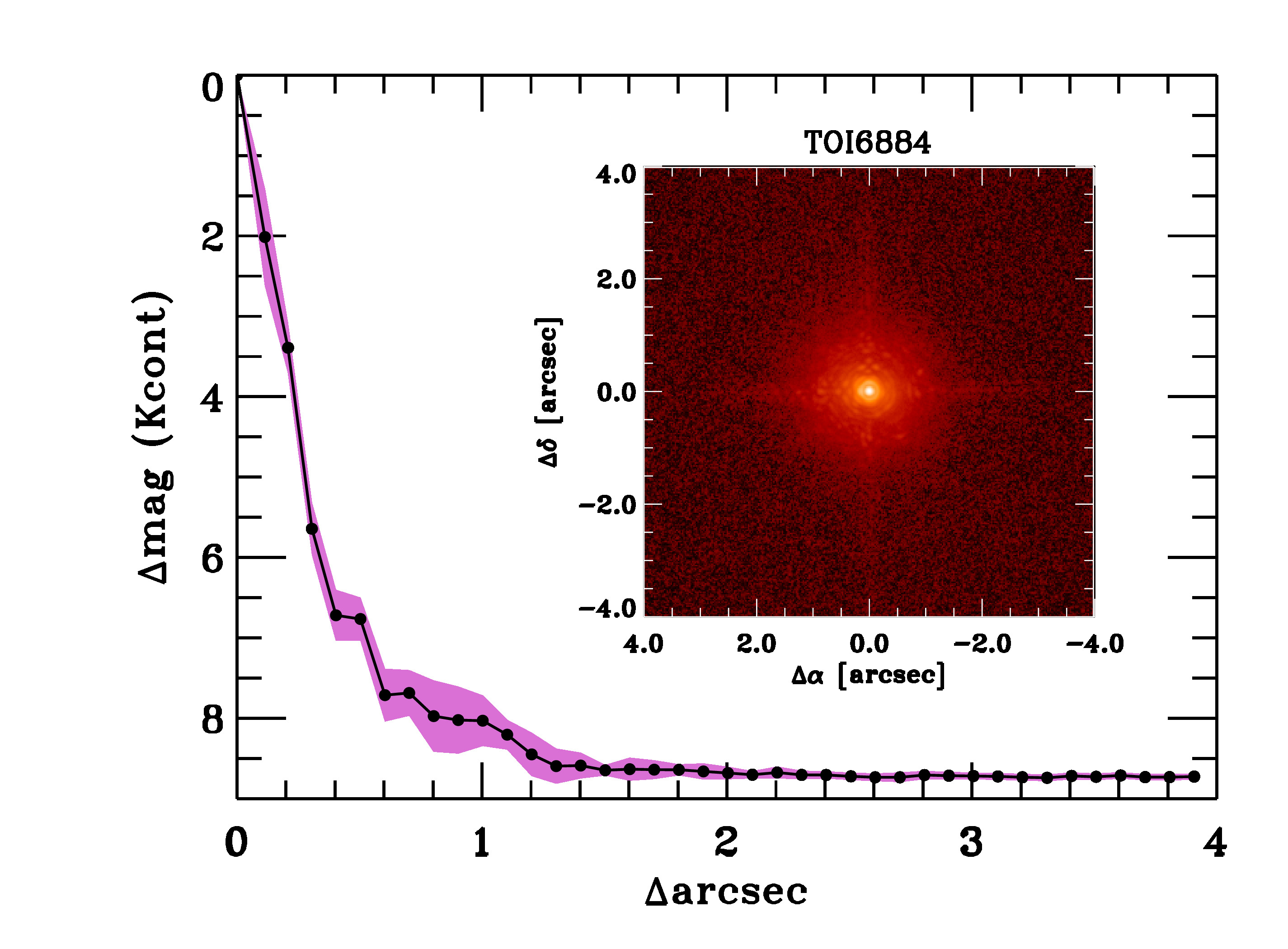}
    \caption{Palomar near-infrared AO imaging and sensitivity curves for TOI-6884 taken in the $K_{cont}$ filter. Inset: Image of the central portion of the data, centered on the star.}
    \label{fig:toi6884_palomar}
\end{figure}

TOI-6884 was observed on 2024 February 15 using the Palomar High Angular Resolution Observer (PHARO) instrument \citep{hayward2001} mounted on the 5\,m Hale Telescope at Palomar Observatory. Observations were carried out behind the natural guide star P3K AO system \citep{dekany2013} using the narrowband $K_{cont}$ filter ($\lambda_0 = 2.26\,\mu$m, $\Delta \lambda = 0.06\,\mu$m). PHARO provides a pixel scale of 0.025\arcsec\ per pixel. The observations consisted of images obtained in sets of 15 following a standard five-point quincunx dither pattern with 5\arcsec\ offsets, repeated three times, with each repeat separated by 0.5\arcsec. The reduced science frames were combined to produce a single mosaicked image with a final angular resolution of 0.10\arcsec. The sensitivity of the final combined AO image was assessed by injecting simulated point sources azimuthally around the primary star at position angles spaced every $20^\circ$, and at separations corresponding to integer multiples of the FWHM of the central source, following the methodology of \citet{Furlan2017_ciardi}. The flux of each injected source was scaled until it was detected at a 5$\sigma$ significance level using standard aperture photometry. At each separation, the final 5$\sigma$ contrast limit was determined as the average of the recovered limits across all azimuthal angles, with uncertainties calculated from the rms dispersion of the azimuthal measurements. The resulting contrast curve is shown in Fig.~\ref{fig:toi6884_palomar}. No companion was detected around TOI-6884 down to a contrast of $\Delta K = 8.0$\,mag relative to the primary star at separations out to 1.0\arcsec.

\subsubsection{SAI observations}
TOI-6884 was observed on UT 2024 February 27 with the speckle polarimeter on the 2.5\,m telescope at the Caucasian Observatory of Sternberg Astronomical Institute (SAI) of Lomonosov Moscow State University. A low--noise CMOS detector Hamamatsu ORCA--quest \citep{Strakhov2023} was used as a detector. The atmospheric dispersion compensator was active, which allowed using the $I_\mathrm{c}$ band. The respective angular resolution is 0.083\arcsec. A total of 6292 frames with 19 ms exposure have been accumulated. The long-exposure FWHM was 0.81\arcsec. We did not detect any stellar companions, detection limits are $\Delta I_\mathrm{c}=3.8^m$ and $5.5^m$ at distances $0.25$ and $1.0^{\prime\prime}$ from the star, respectively. Fig.~\ref{fig:toi6884_sai} shows our final contrast curve.

\subsubsection{WIYN observations}
We observed TOI-6884 using speckle imaging on UT 2024 February 17 with the NN-EXPLORE Exoplanet Stellar Speckle Imager \citep[NESSI;][]{scott2018} at the WIYN 3.5~m telescope on Kitt Peak.  NESSI obtains simultaneous speckle imaging in two filters.  In this case, the filters had central wavelengths $\lambda_c = 562$ and 832~nm. Our observation consisted of a set of 9 1000-frame 40~ms exposures.  The field-of-view was limited by reading out a $256\times256$ pixel subsection of the CCDs, resulting in $4.6\times4.6$~arcsecond images.  Our speckle measurements were further restricted to an outer radius of 1.2~arcseconds from the target star, an area inside of which the speckle patterns were correlated.  A similar observation of a point source standard star was taken in conjunction with that of the TOI.   The standard observation consisted of a single 1000-frame image set and was used to calibrate the intrinsic PSF.
The speckle data were reduced using a pipeline process documented by \cite{Howell2011}.  Among the pipeline products are reconstructed images of the field around TOI-6884 in each filter.  We used these to measure contrast curves, setting detection limits on any point sources close to the TOI (see Fig.~\ref{fig:toi6884_wiyn}).  No companion sources were detected near TOI-6884.


\section{Analysis and Results}\label{sec:analysis}
\subsection{Spectroscopic parameters from spectral synthesis}\label{sec:spec_synthesis}
The TRES spectra were also used to derive spectroscopic parameters of TOI-6884 with the Stellar Parameter Classification tool (SPC; \citealt{Buchhave2012}). SPC cross-correlates each observed spectrum against a grid of synthetic spectra based on the Kurucz atmospheric models \citep{Kurucz1992}. We used all 12 TRES spectra. The average stellar parameters derived with $T_{\rm eff}$, $\log g$, [m/H], and $v\sin i$ as free parameters are: $T_{\rm eff} = 6201 \pm 53$~K, $\log g = 4.16 \pm 0.10$, [m/H] $= 0.05 \pm 0.08$, $v\sin i = 12.9 \pm 0.5$~km~s$^{-1}$, with a typical signal-to-noise per resolution element (SNRe) of 36.1.

\subsection{Stellar Rotation Period and Photometric Modulation}\label{sec:PRot}

\begin{figure}
    \centering
    \includegraphics[width=\columnwidth]{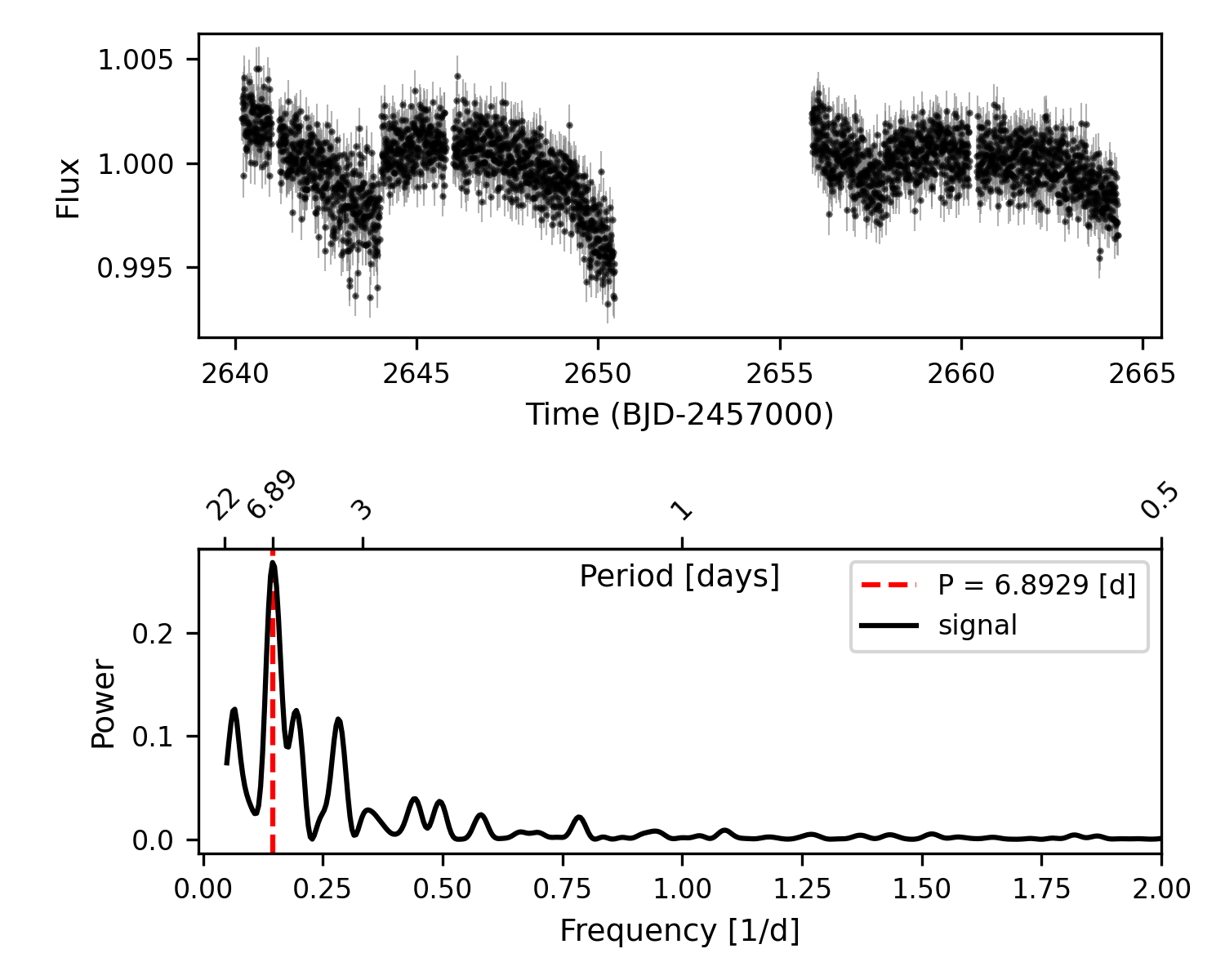}
    \caption{Top panel: \textit{TESS} SAP light curve of TOI-6884, showing the out-of-transit photometric data used for rotation period measurement. Bottom panel: GLS periodogram of the light curve. The vertical dashed red line marks the detected periodic signal at $P = 6.89$ days. The upper axis shows the corresponding period scale.}
    \label{fig:toi6884_tess_sap}
\end{figure}

{We searched for photometric modulation in the \textit{TESS} SAP light curve using out-of-transit data, excluding data with non-zero quality flags as indicated by the SPOC pipeline. We computed a Generalized Lomb--Scargle \citep[GLS;][]{GLS} periodogram, which reveals a periodic signal at $6.89 \pm 0.05$ days (see Fig. \ref{fig:toi6884_tess_sap}). However, this signal is not independently confirmed in data from the SuperWASP all-sky survey \citep[SWASP;][]{Butters2010}.}

TOI-6884 is a moderately rotating star, with a projected rotational velocity of $v \sin i = 12.9 \pm 0.5$ km\,s$^{-1}$ as derived from SPC. The stellar inclination can be estimated by combining the stellar radius, rotation period, and projected rotational velocity. Using the stellar radius ($R_* = 1.840^{+0.072}_{-0.073}\,R_{\odot}$) from our global modeling (Section~\ref{sec:global_modeling}), the photometric rotation period, and the spectroscopic $v \sin i$, and neglecting differential rotation, we estimate a stellar inclination of $72.70 \pm 10.29^\circ$ following \cite{Masuda2020}. The orbital inclination derived from the global modeling is $i = 82.46{^{+0.45}_{-0.42}}^\circ$, suggesting that the system is likely consistent with alignment between the brown dwarf orbital axis and the stellar spin axis.

\subsection{ Galactic Kinematics}
We estimated the Galactic space velocity components ($U$, $V$, $W$) in the barycentric reference frame using the \texttt{gal$\_$uvw}\footnote{\url{https://pyastronomy.readthedocs.io/en/latest/pyaslDoc/aslDoc/gal_uvw.html}} function. The resulting velocities, reported in Table~\ref{tab:star_table}, follow the convention that $U$ is positive toward the Galactic center, $V$ in the direction of Galactic rotation, and $W$ toward the north Galactic pole. We also computed velocities relative to the local standard of rest (LSR) by adopting the solar motion values from \citet{Sch2010}, which are also provided in Table~\ref{tab:star_table}. Based on our analysis, the kinematics of TOI-6884 are consistent with membership in the Galactic thin disk \citep{Leggett1992, Bensby2014}. In addition, we used the BANYAN $\Sigma$ tool, which evaluates membership probabilities based on sky position, proper motion, parallax, and radial velocity, and it classifies TOI-6884 as a field star with a probability greater than 99$\%$, with no association to any known young stellar group \citep{Gagne2018}.

\subsection{Periodogram analysis of RVs}
\begin{figure}
    \centering
    \includegraphics[width=\columnwidth]{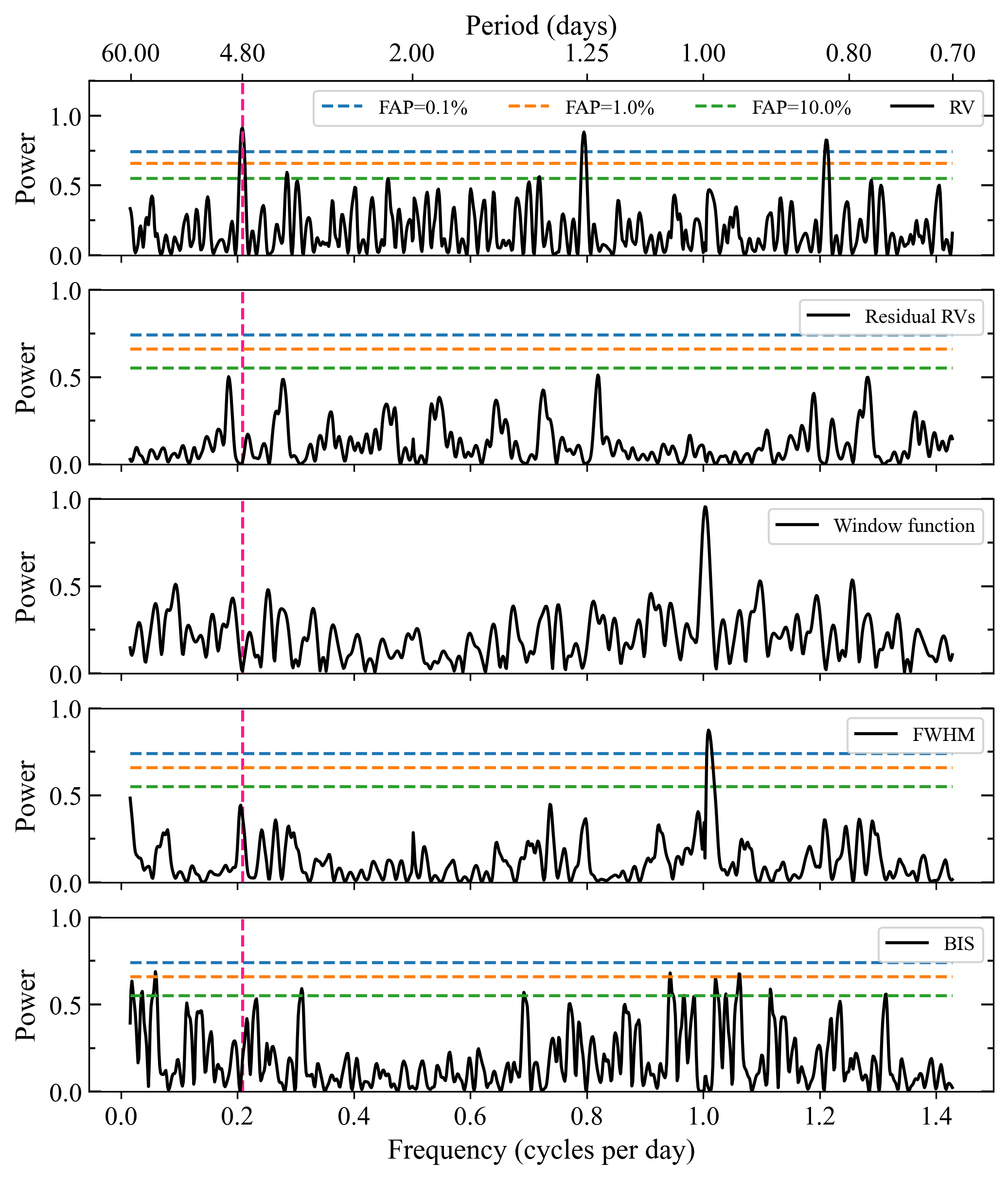}
    \caption{GLS periodograms of TOI-6884. From top to bottom: (1) Radial velocities, (2) Residual RVs after subtracting the dominant signal, (3) Window function, (4) FWHM, and (5) Bisector span. The primary peak at a period of $\approx$ 4.82 days (magenta dashed line) corresponds to the photometrically derived orbital period of the planetary candidate. False alarm probability levels of 0.1$\%$, 1$\%$, and 10$\%$ are indicated as horizontal dashed lines.}
\label{fig:periodogram}
\end{figure}

We computed the GLS for the RVs, residual RVs, window function, FWHM, and bisector span (BIS), shown in panels 1 through 5 (top to bottom) of Fig.~\ref{fig:periodogram}, respectively. The periodogram analysis was performed on RVs obtained from the PARAS-2 spectrograph, following the normalization and false alarm probability (FAP) calculation methods described in \cite{GLS}. Adopting a significance threshold of 0.1\% FAP, we identify a prominent peak at a period of $\approx 4.82 \pm 0.02$ days (indicated by the vertical dashed line in Fig.~\ref{fig:periodogram}), consistent with the orbital period derived from both ground- and space-based transit observations of this planetary candidate.

In addition to the primary planetary signal, the RV periodogram exhibits three other significant peaks at periods of 1.26, 0.82, and 0.56 days. These signals disappear when the $\approx 4.82$-day periodic component is removed using a best-fit sinusoidal model, as evidenced by the clean residual periodogram in panel 2. We attribute these spurious signals to the aliases of the orbital frequency or the 4.82 days signal: the 1/0.82 represents the 1-day alias of $f_{\rm orb}$, whereas, the 1/1.26 and 1/0.56 signals correspond to the 1-day and 2-day aliases of $-f_{\rm orb}$ respectively. No significant peaks above the 0.1\% FAP threshold remain in the residual periodogram after subtracting the main signal.

The spectral window function is shown in panel 3. GLS periodograms of the CCF FWHM (panel 4) and BIS (panel 5) are also shown; these diagnostics quantify line asymmetries that can mimic Doppler shifts and typically serve as indicators of stellar activity. There is no significant signals of the stellar activity in the periodogram is present. Although the CCF FWHM shows a peak coinciding with the spectral window function, it is therefore not considered significant. 

\subsection{Global Fit with EXOFASTv2} \label{sec:global_modeling}

We used EXOFASTv2 \citep[][hereafter EF2]{exofast} to determine both the stellar and planetary parameters of the TOI-6884 system. EF2 is an exoplanet-fitting package written in {\tt IDL} that can simultaneously model both RV and transit data. Additionally, it provides stellar parameters by fitting the spectral energy distribution \citep[SED;][]{SED1} and using MIST stellar evolutionary models \citep{mist_choi, mist_dotter}. The EF2 package utilizes the Markov Chain Monte Carlo (MCMC) algorithm, and convergence is achieved when the Gelman-Rubin statistic is less than 1.01 and the number of independent draws exceeds 1000 \citep{Gelman1992, Gelman_rubin2006}.
\begin{figure*} 
    \centering
    \begin{subfigure}[b]{1.0\columnwidth} 
        \includegraphics[width=\linewidth]{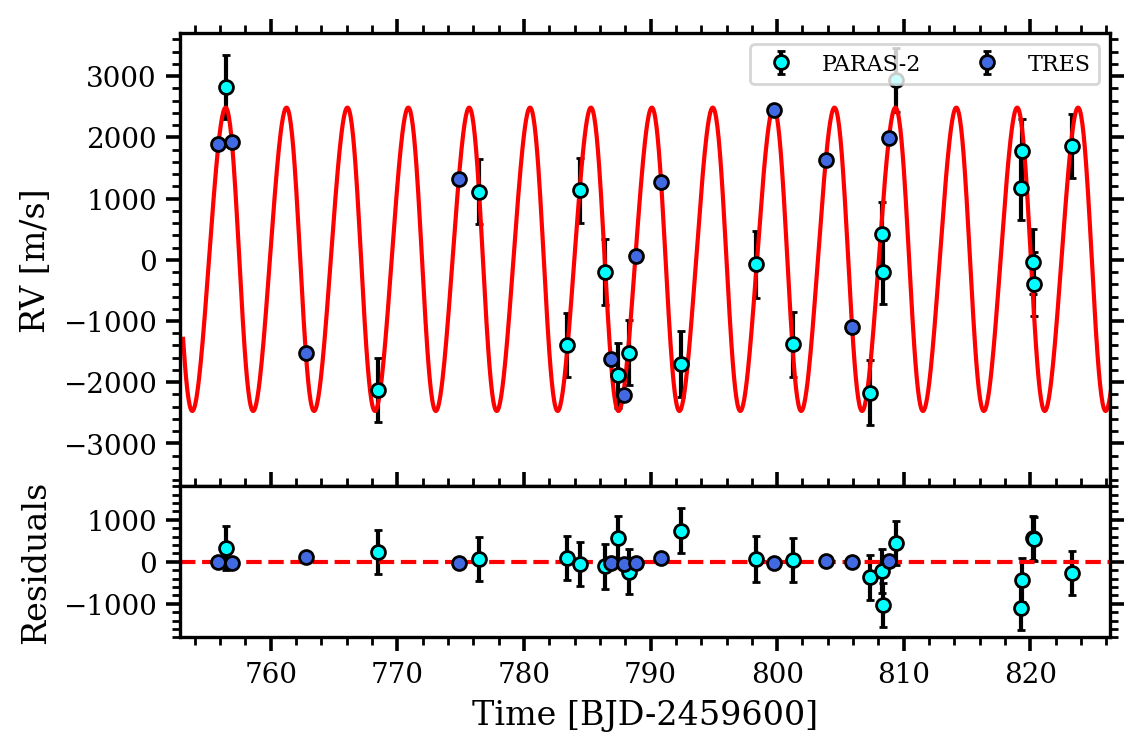}
        \label{fig:RV_full}
    \end{subfigure}
    \begin{subfigure}[b]{1.0\columnwidth} 
        \includegraphics[width=\linewidth]{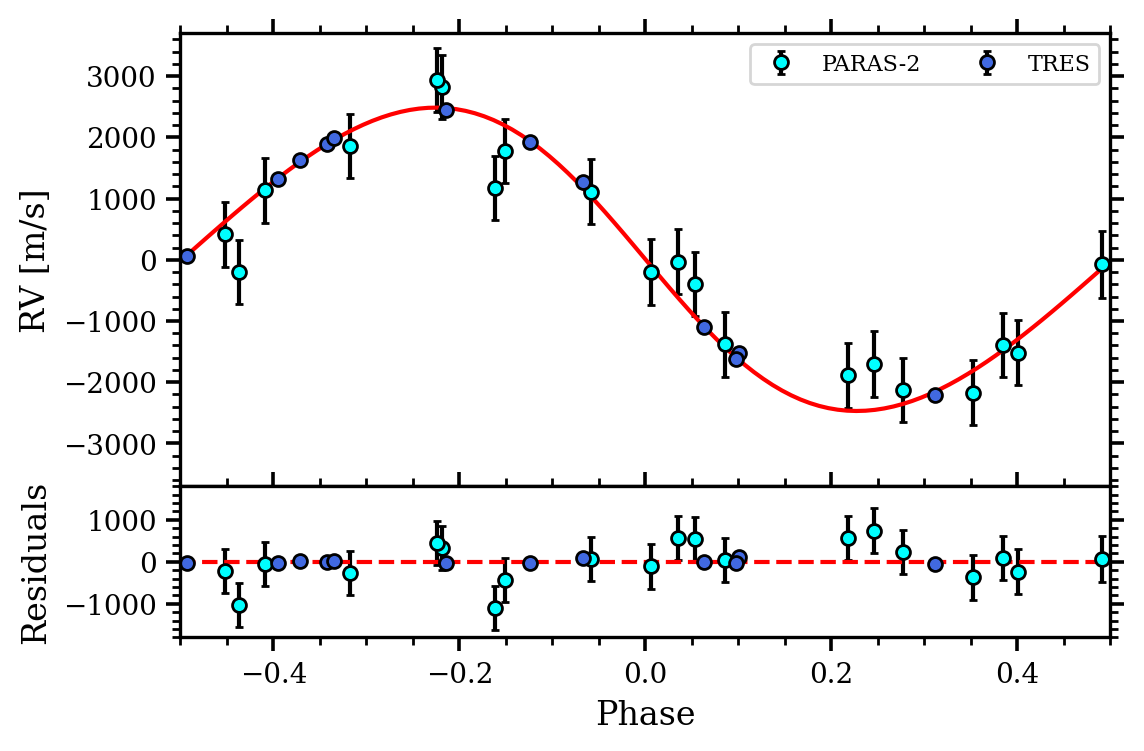}
        \label{fig:RV_folded}
    \end{subfigure}
    \caption{RV measurements of TOI-6884 from PARAS-2 (cyan points) and TRES (blue points) are shown as a function of time (left panel). The phase-folded RVs are shown in the right panel. The red line represents the best-fit RV model using EF2, with the residuals displayed in the lower panels.}
    \label{fig:combined_figures}
\end{figure*}
\begin{figure}
    \centering
    \includegraphics[width=\columnwidth]{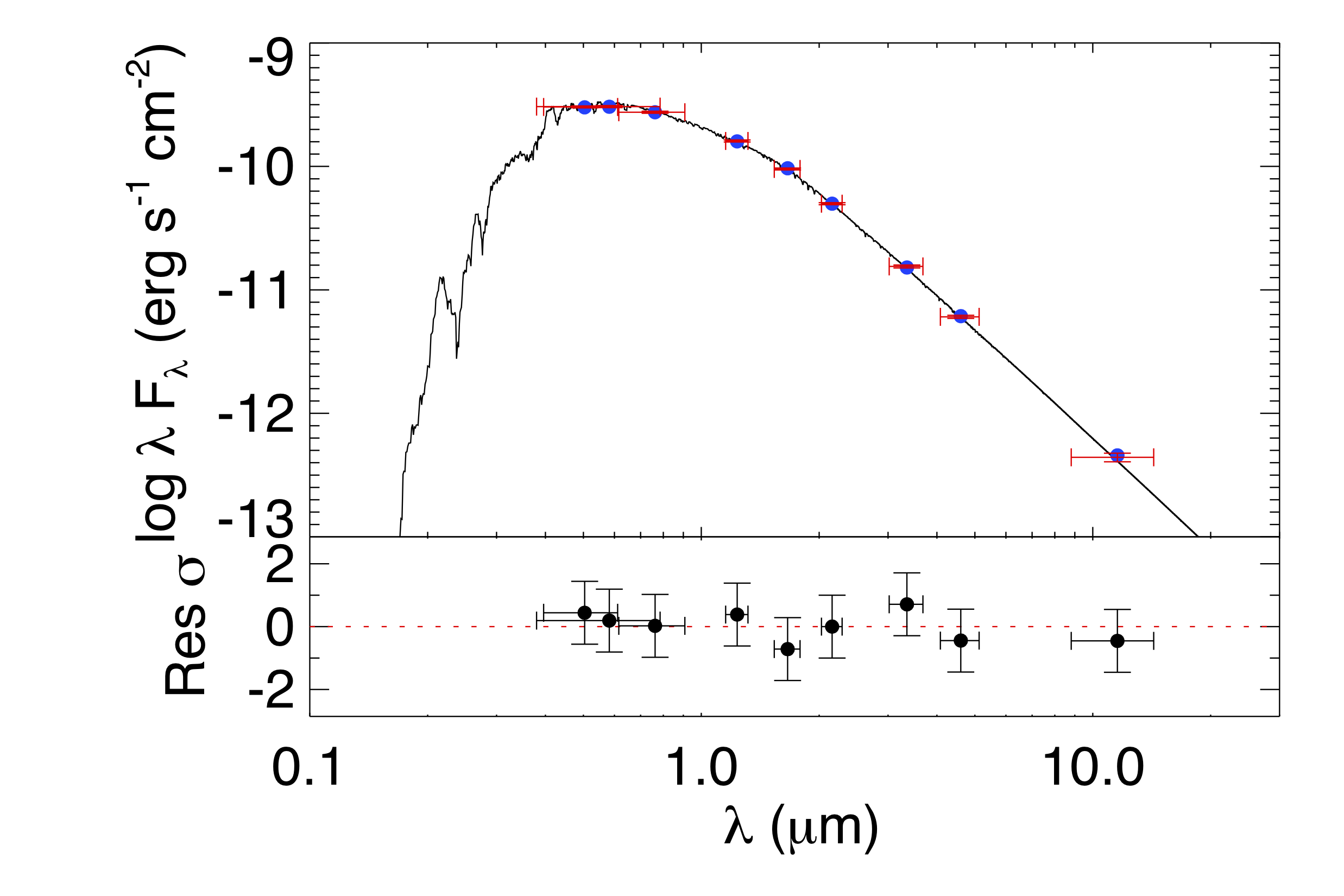}
    \caption{SED of TOI-6884, with red symbols representing the observed photometric measurements and horizontal bars indicating the effective width of the passbands. The blue points represent the model fluxes, and the residuals are displayed in the lower panel.}

    \label{fig:sed}
\end{figure}

\begin{figure}
    \centering
    \includegraphics[width=\columnwidth]{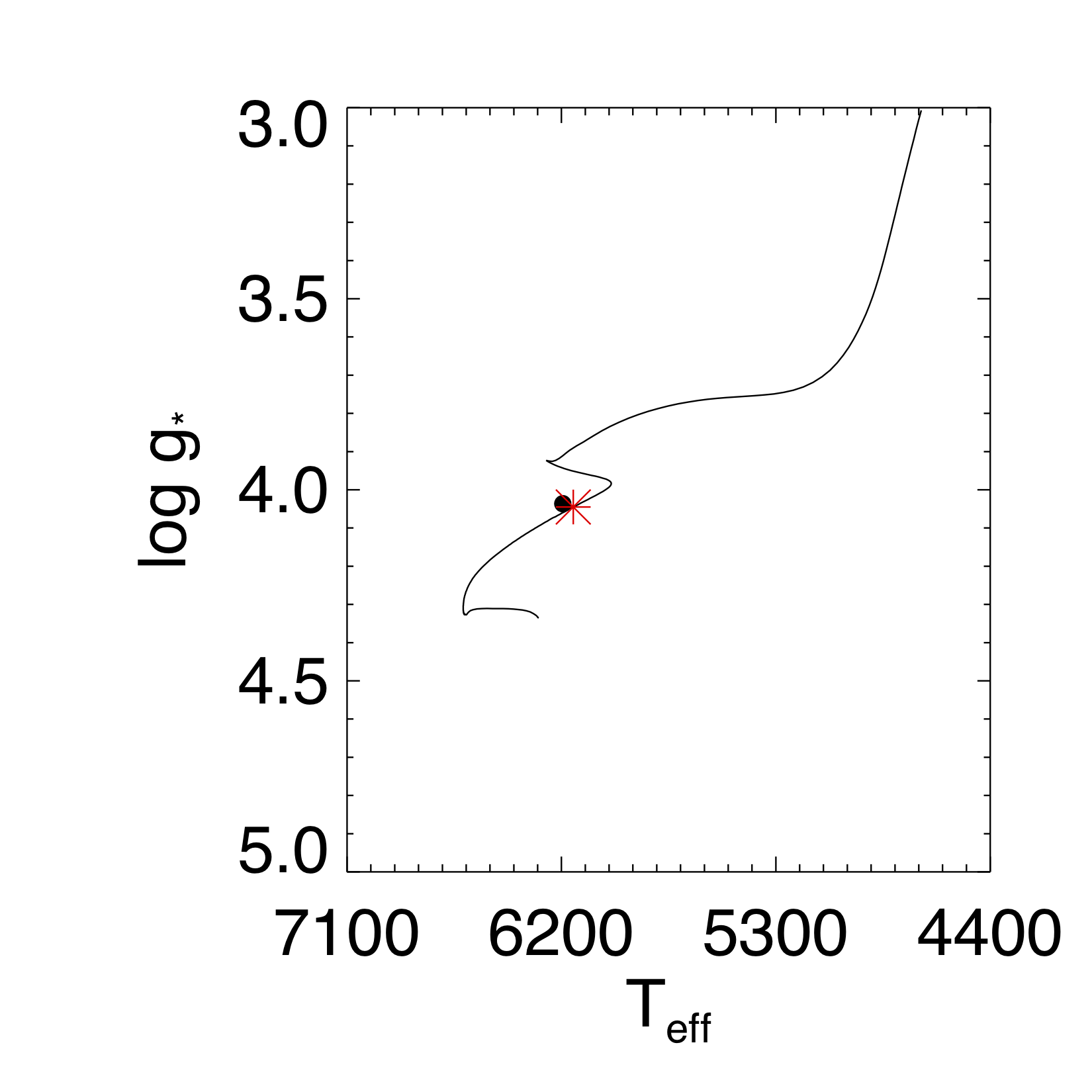}
    \caption{MIST evolutionary track for TOI-6884 shown as a solid black line. The black point indicates the $T_\mathrm{eff}$ and $\log{g}$, while the red asterisk denotes the current age of TOI-6884}
    \label{fig:mist}
\end{figure}

The combination of SED, isochrones, and transit data allows a precise determination of the stellar mass, radius, and surface gravity \citep{Torres2008}. While fitting the RV and transit datasets and modelling the system, we applied Gaussian priors $[\text{Fe}/\text{H}]$ based on the spectroscopic parameters, and on the parallax derived from \textit{Gaia} DR3 \citep{gaia2023} after applying systematic corrections following \cite{Lindegren_2021}. Starting values of $T_{\text{eff}}$ and $\log{g}$ were also provided. For SED fitting, we used photometric magnitudes from the \textit{Gaia} $G$, $G_{BP}$, and $G_{RP}$ bands, along with the 2MASS $J$, $H$, and $K_S$ bands, and the WISE $W1$, $W2$, and $W3$ bands, as listed in Table \ref{tab:star_table}. To account for systematic errors in the absolute photometry, the uncertainties in these magnitudes were inflated following the method outlined in \citet{Lindegren_2021}. Additionally, we applied a uniform prior to enforce an upper limit on the V-band extinction, using the \cite{extinction} dust maps at the location of TOI-6884. 

We performed a global model fit with EF2, keeping the eccentricity as a free parameter. In the process, we also provided the initial values for the orbital period ($P$) and central transit time ($T_C$), which were taken from the ExoFOP website. The priors are listed in Table~\ref{tab:final_param}. A Keplerian orbit is fitted for the RV, while the transit model is created using the approach of \cite{Mandel2002} and \cite{Agol_2020}, with limb-darkening parameters constrained by \cite{Claret} and \cite{Claret_tess}. The EF2 fits revealed slight bimodality in the posterior distributions of stellar mass and age. This bimodality is not unprecedented; it is a known feature in certain regions of the MIST models \citep[e.g.][]{toi1789, Baliwal2024, toi6038}. The two peaks in mass are found at 1.21 and 1.41\msun. We split the distribution at the local minimum of 1.27{\msun} and extracted both solutions, ultimately adopting the high-mass solution due to its higher relative probability (77\%). All reported parameters and subsequent analyses are based on this solution.

The best-fit Kurucz stellar atmosphere model from the SED and the best-fit MIST stellar evolutionary model are shown in Fig. \ref{fig:sed} and Fig. \ref{fig:mist}, respectively. We find that the host star is a late F-type slightly evolved star with parameters $M_* = 1.410^{+0.075}_{-0.069}\ \msun$, $R_* = 1.840^{+0.072}_{-0.073} \ \rsun$, $T_{\mathrm{eff}} = 6330^{+180}_{-160}$ K, $\log{g} = 4.057^{+0.045}_{-0.039}$, and an age of $2.61^{+0.78}_{-0.75}$ Gyr. These parameters are consistent with those derived from spectral synthesis in Sec. \ref{sec:spec_synthesis}. The best-fit transit light curve and RV models are shown in Fig.~\ref{fig:toi6884_transit}, \ref{fig:ground_based_transit} and Fig.~\ref{fig:combined_figures}, respectively. Joint modeling reveals a BD companion with a radius $R_P = 0.927^{+0.051}_{-0.052}\ R_{J}$ and a mass $M_P = 26.32^{+0.98}_{-0.93}\ M_{J}$ in a nearly circular orbit $e=0.067^{+0.010}_{-0.012}$ with an orbital period of $P = 4.808264^{+0.000015}_{-0.000014}$ days. We also calculate a Lucy--Sweeney probability \citep{Lucy1971} of $2.96\times10^{-5}\%$, confirming that the measured eccentricity is highly significant and not spurious. The best-fit stellar and planetary parameters, along with their 68\% confidence intervals for the high-mass solution and low mass solutions, are listed in Table \ref{tab:final_param} and \ref{tab:final_param1}.

\section{Discussion}\label{sec:discussion}
\subsection{The transiting BD population and the Mass–Radius relation}
\begin{figure}
    \centering
    \includegraphics[width=\columnwidth]{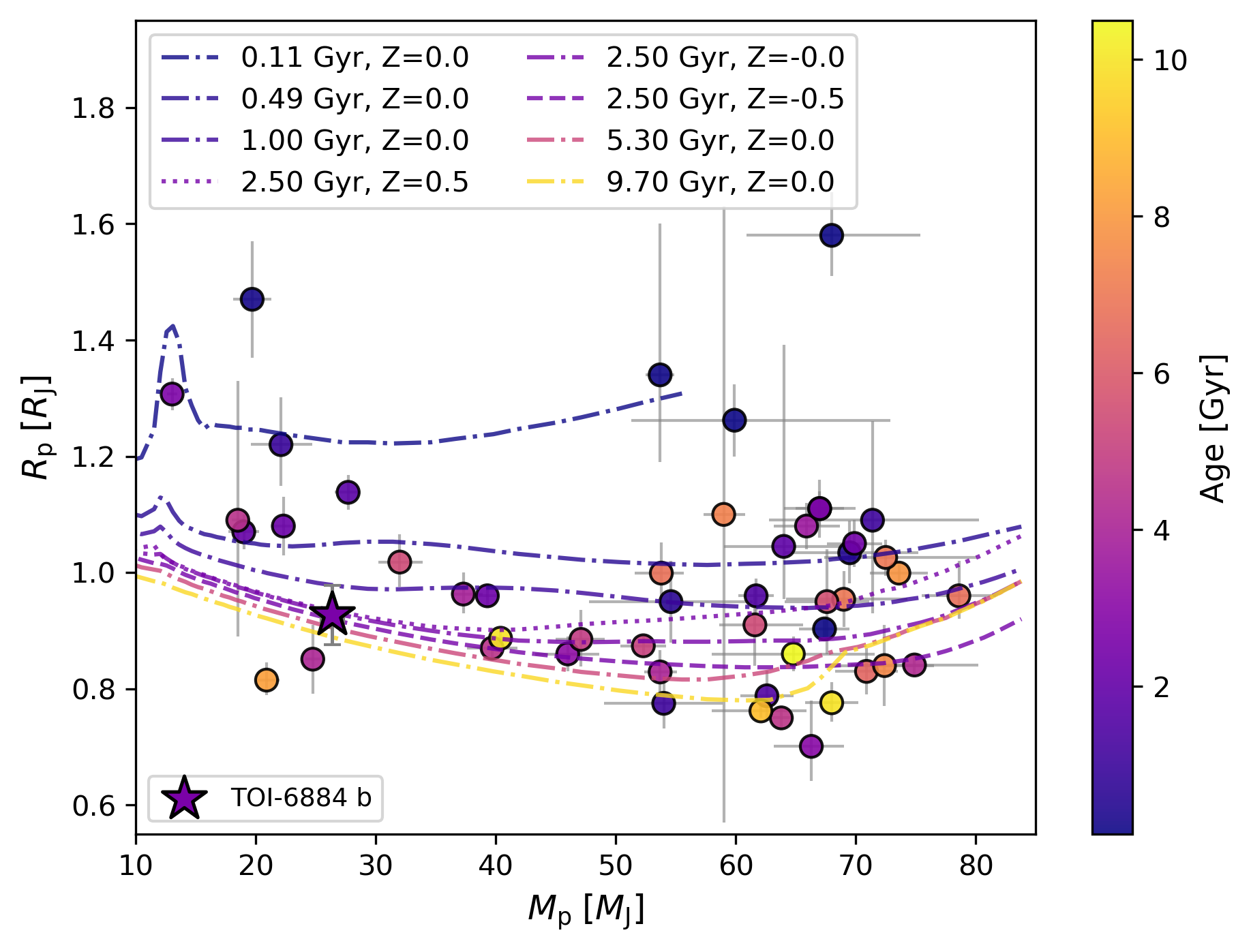}
    \caption{$M-R$ diagram of all known transiting BDs with masses between 13\mj~and 80\mj. TOI-6884b is marked with a star. Points are color-coded by the published ages of their host systems. Colored lines show evolutionary isochrones from \citet{Morley2024} at ages of 0.11, 0.49, 1.0, 2.5, 5.2, and 9.3~Gyr.  Different line styles indicate metallicity variations ([M/H] = –0.5, 0.0, +0.5) for the 2.5~Gyr isochrone.}

    \label{fig:mr_plot}
\end{figure}

We compiled a comparison sample of transiting BDs and gathered their properties from the TEPCat catalogue \citep[][and references therein]{tepcat}. For this discussion, we define a brown dwarf as a substellar object with a mass between 13\mj~and 80\mj. There are a total of 49 transiting BDs in the sample and TOI-6884b adds one new member to this population, with a precisely determined mass and radius (i.e., with <10$\%$ uncertainties).
Fig.~\ref{fig:mr_plot} presents the mass–radius ($M-R$) distribution for these systems, including TOI-6884b. The BDs are color-coded by the ages of their host stars, and we overplot evolutionary models isochrones from the Sonora Diamondback grid \citep[S24;][]{Morley2024}, using the same color coding as the BD sample. We chose the S24 models because they account for the effects of both clouds and metallicity. In particular, we adopt the `hybrid' case where clouds are included above 1300 K and cloud-free below 1300 K. We show substellar models for ages of 0.11 Gyr, 0.49 Gyr, 1 Gyr, 2.5 Gyr, 5.3 Gyr, and 9.7 Gyr in the figure. To explore metallicity effects, we also include the 2.5~Gyr isochrone for [M/H] of –0.5 and +0.5. 

Our analysis yields a mass and radius for TOI-6884b of $26.32^{+0.98}_{-0.93}~M_{\rm J}$ and $0.927^{+0.051}_{-0.052}~R_{\rm J}$, respectively. The stellar age derived from MIST isochrones shows a bimodal solution of either 2.61 or 5.19~Gyr, with respective probabilities of 0.77 and 0.23. As shown in Figure~\ref{fig:mr_plot}, TOI-6884b lies between the 1~Gyr and 9.7~Gyr isochrones, consistent with an age in the 2.6--5.2~Gyr range indicated by both the S24 and MIST models. This agreement remains robust even when accounting for metallicity variations, reinforcing the reliability of the inferred age.

Moreover, to investigate how incident stellar flux influences the radii of hot Jupiters, BDs, and low-mass stars, \citet{Lin2023} derived a flux–radius power-law relation for BDs and low-mass stars, analogous to the relation previously obtained by \citet{Weiss2013} for hot Jupiters. Their analysis yielded an empirical flux–radius relation for BDs in the mass range 13–80~$M_{\rm J}$:
\begin{equation}\label{eq:lin2023}
\frac{R_{\rm p}}{R_{\rm J}} = 1.11 \left( \frac{M_{\rm p}}{M_{\rm J}} \right)^{-0.052} \left( \frac{F}{F_\oplus} \right)^{0.009}.
\end{equation}
{For TOI-6884b, the predicted radius from this relation, $R_{\rm p} = 0.998\,R_{\rm J}$ (Equation~\ref{eq:lin2023}), agrees well within uncertainties with our radius measurement, showing no evidence of radius inflation. We also consider the substellar evolutionary models of \citet{Mukherjee2025}, which explicitly incorporate the effects of incident stellar flux (expressed as $\log F$) on the radii of irradiated brown dwarfs as a function of mass and age. These models predict that strong irradiation (e.g., $\log_{10}(F/{\rm cgs}) \gtrsim 9$, corresponding to $T_{\rm eq} \gtrsim 1450$ K) can lead to radius enhancement, particularly at younger ages. TOI-6884b receives an incident flux of $1.68^{+0.14}_{-0.13} \times 10^{9}$~cgs, placing it within this irradiated regime. However, \citet{Mukherjee2025} also show that the impact of irradiation on brown dwarf radii decreases significantly at Gyr timescales (see Fig. 4; \citealp{Mukherjee2025}). In this context, the radius of TOI-6884b remains consistent with the empirical flux–radius relation of \citet{Lin2023}, suggesting no strong evidence for anomalous inflation. This interpretation is further supported by theoretical studies on giant planets \citep[e.g.,][]{Burrows2007}, which indicate that the effect of stellar irradiation on planet radii diminishes with increasing object mass at an age of $\sim$2.5 Gyr. 
These comparisons suggest that TOI-6884b does not exhibit significant radius inflation despite its relatively high incident flux, consistent with earlier suggestions that irradiation has a weaker impact on brown dwarfs than on hot Jupiters \citep{Bouchy2011}. Nevertheless, the non-zero flux dependence in the \citet{Lin2023} relation indicates that irradiation effects may not be entirely negligible \citep[e.g.,][]{Page2024,Psaridi2022}. A larger and uniformly characterized sample of transiting BDs will be essential to better constrain the role of irradiation in shaping their radii.}

\subsection{Tidal evolution}\label{sec:tidal_evolution}
Given the small orbital separation of the TOI-6884 system, tidal effects play a dominant role in its orbital evolution and circularisation. At present, there is no clear understanding of how to determine tidal quality factors for a system without detailed knowledge of the composition of the bodies \citep[e.g.,][]{Jackson+2008}. Therefore, we adopt a broad range of tidal quality factor values from the literature and estimate the circularisation timescales using the equations described in \citet{Jackson+2009} and their approximations as discussed in \cite{Henderson2024}:
\begin{equation}\label{eq:tidal1}
    \frac{1}{a}\frac{\mathrm da}{\mathrm dt}=-\left[\frac{63}{2}\left(\frac{R_p^5\sqrt{GM_\star^3}}{Q_p'M_p}\right)e^2+
    \frac{9}{2}\sqrt{\frac{G}{M_\star}}\frac{R_\star^5M_p}{Q_\star'}\left(1+\frac{57}{4}e^2\right)\right]a^{-\frac{13}{2}}
\end{equation}
\begin{equation}\label{eq:tidal2}
    \frac{1}{e}\frac{\mathrm de}{\mathrm dt}=-\left[\frac{63}{4}\sqrt{GM_\star^3}\left(\frac{R_p^5}{Q_p'M_p}\right)+\frac{225}{16}\sqrt{\frac{G}{M_\star}}\left(\frac{R_\star^5M_p}{Q_\star'}\right)\right]a^{-\frac{13}{2}} 
\end{equation}
\begin{figure*}
    \centering
    \includegraphics[width=\linewidth]{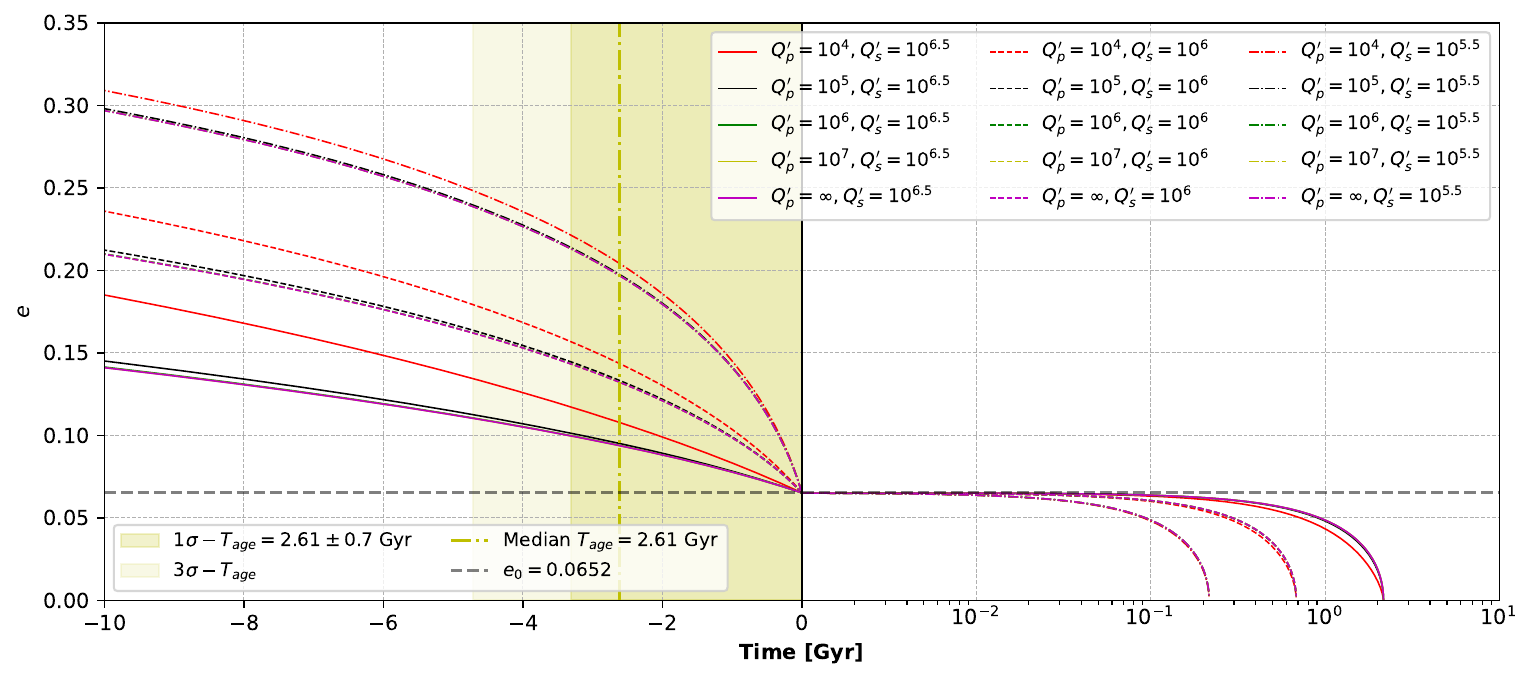}
    \caption{Backward and forward evolution of the orbital separation for different values of $Q_\star'$ and $Q_p'$, based on the formulation of \citet{Jackson+2009}, for the high-probability solution of the TOI-6884 system. The low-probability solution is shown in Fig~\ref{fig:e-Qs-full_low}.}
    \label{fig:e-Qs-full}
\end{figure*}
The results are summarized in Table \ref{tab:tidal_combined}. In planetary systems, the effects of stellar tides (represented by $Q_{\star}'$) are often neglected compared to planetary tides (represented by $Q_p'$) because of the small mass ratios. This approximation can be considered to provide an upper limit on the circularisation timescales. Given the nearly circularized orbit of TOI-6884b ($e \sim 0.06$) and the age of the system ($\approx$ 2.61 Gyr), we can put the constraints $Q_p' \gtrsim 10^5$ and $Q_\star' \lesssim 10^{7.5}$, purely based on the argument that the system's circularisation should happen within its lifetime (see Table \ref{tab:tidal_combined}). These estimates are also in broad agreement with other conservative estimates adopted in the literature \citep{Lanza+2011, Beatty2018}. 


The prescription used above inherently assumes that the variation in the semi-major axis is negligible as the circularisation happens. This need not be the case if stellar tides are significant for a system \citep{Jackson+2008}. Therefore, in order to evaluate the impact of stellar tides on these estimates, we performed a complete numerical integration of equations \ref{eq:tidal1} and \ref{eq:tidal2} over a forward and backward period of 10 Gyr. {It should be noted that higher order effects start coming up for $a\gtrsim0.2\,\text{AU}$ in which case the validity of the equations \ref{eq:tidal1} and \ref{eq:tidal2} becomes uncertain (see discussion in \cite{Jackson+2008, Jackson+2009}).} We considered different values of $Q_p'$ for specific representative values of $Q_\star'$ to assess their effect on the eccentricity evolution of the system and the results are presented in Fig. \ref{fig:e-Qs-full} and \ref{fig:e-Qs-full_low}. It is evident that stronger stellar tides lead to faster circularisation (dash-dotted curves compared to solid ones). This behaviour seems to be in broad agreement with the conservative approach used to calculate the approximate circularisation timescales, thereby putting more confidence on the tidal parameter constraints derived therein.

{We estimated the tidal synchronization timescales following the formalism of \citet{Albrecht2012}, obtaining $\tau_{\rm CE} = 1.36^{+0.35}_{-0.33}\,\mathrm{Gyr}$ and $\tau_{\rm RA} = 2.55^{\,+0.91}_{\, -0.78}\times10^{4}\,\mathrm{Gyr}$. Here, $\tau_{\rm CE}$ corresponds to stars with convective envelopes (typically $T_{\rm eff} \lesssim 6250\,\mathrm{K}$), while $\tau_{\rm RA}$ applies to stars with radiative envelopes ($T_{\rm eff} \gtrsim 6250\,\mathrm{K}$), where tidal dissipation is significantly less efficient.
The transition between these two regimes occurs near the Kraft break \citep{Kraft1967}, at $T_{\rm eff} \sim 6250\,\mathrm{K}$, marking a change in both tidal dissipation efficiency and magnetic braking. With an effective temperature of $T_{\rm eff} = 6330^{+180}_{-160}\,\mathrm{K}$, the host star lies close to this boundary. If the star possesses a radiative envelope, the synchronization timescale greatly exceeds the system age, implying inefficient tidal synchronization. In contrast, assuming a convective envelope yields a timescale shorter than the system age, suggesting that the system should be tidally synchronized. However, the observed stellar rotation period, $P_{\rm rot} = 6.89 \pm 0.05\,\mathrm{d}$, differs from the orbital period, $P_{\rm orb} = 4.808\,\mathrm{d}$, indicating that the system is not synchronized. This behavior is more consistent with the radiative-envelope regime and suggests inefficient tidal dissipation. The discrepancy between theoretical expectations and observations may reflect uncertainties in tidal dissipation mechanisms, making this system a compelling target for further investigation.}
\begin{table}
\caption{Tidal circularisation timescales for the TOI-6884 system estimated using approximations to tidal evolution equations \citep{Jackson+2008, Jackson+2009} for both high and low probability mass solutions derived from EF2.}
\label{tab:tidal_combined}
\centering
\begin{tabular}{llcc}
\hline\hline
\noalign{\smallskip}
$Q_\star'$ & $Q_p'$ & \multicolumn{2}{c}{$\tau_{\text{circ}}^{\text{eff}}$ (Gyr)} \\
\noalign{\smallskip}
            &          & High-probability & Low-probability \\
\noalign{\smallskip}
\hline
\noalign{\smallskip}
$10^{5}$   & $10^{4.5}$ & 0.02 & 0.02 \\
           & $10^{5}$   & 0.02 & 0.02 \\
           & $10^{5.5}$ & 0.02 & 0.02 \\
           & $10^{6}$   & 0.02 & 0.02 \\
           & $10^{6.5}$ & 0.02 & 0.02 \\
\noalign{\smallskip}\hline
\noalign{\smallskip}
$10^{5.5}$ & $10^{4.5}$ & 0.07 & 0.05 \\
           & $10^{5}$   & 0.07 & 0.05 \\
           & $10^{5.5}$ & 0.07 & 0.05 \\
           & $10^{6}$   & 0.07 & 0.05 \\
           & $10^{6.5}$ & 0.07 & 0.05 \\
\noalign{\smallskip}\hline
\noalign{\smallskip}
$10^{6}$   & $10^{4.5}$ & 0.23 & 0.16 \\
           & $10^{5}$   & 0.23 & 0.16 \\
           & $10^{5.5}$ & 0.24 & 0.16 \\
           & $10^{6}$   & 0.24 & 0.16 \\
           & $10^{6.5}$ & 0.24 & 0.16 \\
\noalign{\smallskip}\hline
\noalign{\smallskip}
$10^{7}$   & $10^{4.5}$ & 1.60 & 1.10 \\
           & $10^{5}$   & 2.05 & 1.42 \\
           & $10^{5.5}$ & 2.26 & 1.56 \\
           & $10^{6}$   & 2.33 & 1.61 \\
           & $10^{6.5}$ & 2.36 & 1.62 \\
\noalign{\smallskip}\hline
\noalign{\smallskip}
$10^{7.5}$   & $10^{4.5}$ & 2.96 & 2.05 \\
           & $10^{5}$   & 5.04 & 3.48 \\
           & $10^{5.5}$ & 6.49 & 4.48 \\
           & $10^{6}$   & 7.14 & 4.92 \\
           & $10^{6.5}$ & 7.37 & 5.08 \\
\noalign{\smallskip}\hline
\noalign{\smallskip}
$Q_{\star}' \to \infty$ & $10^{4.5}$ & 4.89 & 3.39 \\
                        & $10^{5}$   & 15.46 & 10.71 \\
                        & $10^{5.5}$ & 48.90 & 33.88 \\
                        & $10^{6}$   & 154.63 & 107.14 \\
                        & $10^{6.5}$ & 488.98 & 338.80 \\
\noalign{\smallskip}\hline
\end{tabular}
\end{table}

\subsection{Eccentricity distribution}
The orbital period and eccentricity together provide valuable insights into the dynamical evolution of transiting systems. As discussed in Section~\ref{sec:tidal_evolution}, the low eccentricity of TOI-6884b is consistent with its relatively short tidal circularisation timescale. {The low eccentricity of TOI-6884 b ($e = 0.067^{+0.010}_{-0.012}$) is consistent with the ``planet-like'' distribution often observed for BDs below 42.5 \mj \citep{Ma2014}. Though this behavioral transition remains a subject of debate \citep{Vowell2025}.}
{Moreover, in the context of giant planets, \citet{Grunblatt2018} found that those orbiting evolved stars, primarily red giant branch (RGB) hosts, tend to exhibit higher eccentricities (median $e \approx 0.152$) than planets around main-sequence hosts (median $e \approx 0.056$). This trend has been attributed to enhanced tidal interactions as stellar radii expand during post-main-sequence evolution. However, such systems are not directly comparable to subgiant hosts, which have not yet undergone the substantial radius expansion characteristic of red giants. In these systems, planets with shorter tidal circularisation timescales are expected to undergo rapid orbital decay and eventual engulfment, leaving behind a population dominated by high-eccentricity systems. Extending this trend to BDs orbiting subgiant stars is therefore not straightforward. In addition, the higher masses of BDs compared to giant planets can lead to stronger tidal interactions and shorter circularisation timescales, potentially altering their orbital evolution. In this context, TOI-6884b exhibits a relatively low eccentricity ($e=0.067^{+0.010}_{-0.012}$), which may be consistent with efficient tidal circularisation. Currently, only 11 BDs have been detected around evolved stars, including TOI-6884b. Among these, only two, TOI-5882b \citep{Vowell2025} and TOI-1994b \citep{Page2024}, fall within the low-mass regime and have masses comparable to TOI-6884b. Given the limited sample size, it remains unclear whether the orbital architectures of BDs around evolved stars systematically differ from those of planets around main-sequence or red giant hosts.}

\subsection{Future work}
TOI-6884\,b presents several opportunities for future follow-up, particularly in the context of atmospheric characterisation and orbital architecture. Using the framework of \citet{Kempton_2018}, we estimate the transmission spectroscopy metric (TSM) and emission spectroscopy metric (ESM) for this system. The derived TSM value of $\sim$0.9 indicates that the object is not a favourable target for transmission spectroscopy. We note, however, that the TSM formalism was primarily developed for smaller exoplanets, and its applicability to objects in the BD regime is uncertain. Similarly, the ESM metric of \citet{Kempton_2018} was formulated for irradiated exoplanets and does not explicitly account for the intrinsic thermal emission of a BD. As a result, our estimated ESM value of $\sim$16 should be regarded as a lower limit, since the BD’s own emission is expected to further enhance the planet-to-star flux contrast at infrared wavelengths. Even under this conservative assumption, the estimated ESM exceeds the threshold value of 7.5 suggested by \citet{Kempton_2018}, indicating that TOI-6884\,b is a promising target for atmospheric characterisation via emission spectroscopy. The system is therefore well suited for detailed atmospheric studies with current and upcoming facilities such as the \textit{James Webb Space Telescope} (JWST; \citealt{Gardner_2006}) and \textit{ARIEL} \citep{Tinetti_2018}.

Beyond atmospheric studies, TOI-6884\,b is also a promising target for investigating its orbital configuration via the Rossiter--McLaughlin \citep[RM;][]{Triaud2018} effect, {which provides a robust measurement of the sky-projected obliquity of the system. If the inclination of the stellar spin axis is known, this measurement can be used to infer the true stellar obliquity. As discussed in Section~\ref{sec:PRot}, TOI-6884\,b is likely to be aligned based on our estimate of the stellar inclination angle; however, the system obliquity remains unconstrained.}
The expected RM semi-amplitude of $\sim 33$\,m\,s$^{-1}$ is well within the capabilities of current high-resolution spectrographs, enabling detection and modelling of the RM anomaly during transit. Such observations would yield a direct measurement of the spin--orbit angle, $\lambda$, providing valuable constraints on the formation and dynamical evolution of this close-in brown dwarf. Combined atmospheric and RM observations would therefore offer a more complete picture of the physical and dynamical properties of the TOI-6884 system.
\section{Summary}\label{sec:summary}

In this work, we report the discovery and characterisation of TOI-6884\,b, a transiting BD orbiting a {slightly} evloved F-type star. The system was initially identified as an exoplanet candidate from transit signals detected by NASA’s \textit{TESS} mission with a reported orbital period of $\sim$14.42~days. Through careful re-analysis of the \textit{TESS} photometry and confirmation with ground-based follow-up observations, we identified the true orbital period to be $4.808264^{+0.000015}_{-0.000014}$~days, corresponding to a one-third harmonic of the initially reported value. The substellar nature of the companion was confirmed through extensive follow-up, including high-precision radial velocity measurements with PARAS-2 at the PRL 2.5\,m telescope and TRES at FLWO, time-series transit photometry obtained with TCS/MuSCAT-2, LCO/Teide, LCO/McDonald, and LCO-HAL/MuSCAT-3, as well as high-angular-resolution imaging from Gemini, WIYN, and SAI, which rule out close stellar companions. From a joint analysis of the photometric and spectroscopic data, we determine that the host star is an F-type star with an effective temperature of $6330^{+180}_{-160}$~K, a metallicity of $[{\rm Fe/H}]=0.094^{+0.073}_{-0.068}$, and stellar parameters consistent with a slightly evolved star. The derived stellar mass, radius, and age are $1.410^{+0.075}_{-0.069}\,M_{\odot}$, $1.840^{+0.072}_{-0.073}\,R_{\odot}$, and $2.61^{+0.78}_{-0.75}$~Gyr, respectively. The companion, TOI-6884\,b, has a mass of $26.32^{+0.98}_{-0.93}\,M_{\rm J}$ and a radius of $0.927^{+0.051}_{-0.052}\,R_{\rm J}$, placing it securely in the BD regime. Its orbit exhibits a small but non-zero eccentricity of $0.067^{+0.010}_{-0.012}$, and the estimated equilibrium temperature is $1652 \pm 32$~K. TOI-6884\,b adds to the small but growing sample of well-characterised transiting BDs, particularly in the regime of short orbital periods around slightly evolved, late-type host stars. Such systems provide important empirical constraints on the mass--radius relation and orbital properties of substellar companions, and contribute to improving our understanding of BD structure and evolution across different irradiation and tidal environments.
\section*{Acknowledgements}

We acknowledge the PRL-DOS (Department of Space, Government of India) and the director of PRL for supporting the PARAS spectrograph funding for the exoplanet discovery project and the research grant for AK, SB, and SD. We also acknowledge support from the Swiss National Science Foundation IZSTZ0$\_$216537 and the Centre for Space and Habitability (CSH) of the University of Bern. Part of this work received support from the National Centre for Competence in Research PlanetS, supported by the Swiss National Science Foundation (SNSF). YGMC, MPM, SP, and AK are partially supported
by UNAM PAPIIT-IG101224. The work of B.S. and P.B. was conducted under the state assignment of Lomonosov Moscow State University. Participation of KN was made possible by the SETI Institute REU internship program (NSF award 2051007). This work makes use of observations from the LCOGT network. Part of the LCOGT telescope time was granted by NOIRLab through the Mid-Scale Innovations Program (MSIP). MSIP is funded by NSF. This paper is based on observations made with the Las Cumbres Observatory’s education network telescopes that were upgraded through generous support from the Gordon and Betty Moore Foundation. This article is based on observations made with the MuSCAT2 instrument, developed by ABC, at Telescopio Carlos Sánchez operated on the island of Tenerife by the IAC in the Spanish Observatorio del Teide. This paper is based on observations made with the MuSCAT3 instrument, developed by the Astrobiology Center and under financial supports by JSPS KAKENHI (JP18H05439) and JST PRESTO (JPMJPR1775), at Faulkes Telescope North on Maui, HI, operated by the Las Cumbres Observatory. This work is partly supported by JSPS KAKENHI Grant Numbers JP24H00017, JP24K00689, JSPS Bilateral Program Number JPJSBP120249910 and JSPS Grant-in-Aid for JSPS Fellows Grant Number JP25KJ0091. This work is partly financed by the Spanish Ministry of Economics and Competitiveness through grants PGC2018-098153-B-C31. This research has made use of the Exoplanet Follow-up Observation Program (ExoFOP; DOI: 10.26134/ExoFOP5) website, which is operated by the California Institute of Technology, under contract with the National Aeronautics and Space Administration under the Exoplanet Exploration Program. Funding for the TESS mission is provided by NASA's Science Mission Directorate. KAC and CNW acknowledge support from the TESS mission via subaward s3449 from MIT. Some of the observations in this paper made use of the High-Resolution Imaging instrument ‘Alopeke and were obtained under Gemini LLP Proposal Number: GN/S-2021A-LP-105. ‘Alopeke was funded by the NASA Exoplanet Exploration Program and built at the NASA Ames Research Center by Steve B. Howell, Nic Scott, Elliott P. Horch, and Emmett Quigley. Alopeke was mounted on the Gemini North telescope of the international Gemini Observatory, a program of NSF’s OIR Lab, which is managed by the Association of Universities for Research in Astronomy (AURA) under a cooperative agreement with the National Science Foundation. on behalf of the Gemini partnership: the National Science Foundation (United States), National Research Council (Canada), Agencia Nacional de Investigación y Desarrollo (Chile), Ministerio de Ciencia, Tecnología e Innovación (Argentina), Ministério da Ciência, Tecnologia, Inovações e Comunicações (Brazil), and Korea Astronomy and Space Science Institute (Republic of Korea). We acknowledge financial support from the Agencia Estatal de Investigaci\'on of the Ministerio de Ciencia e Innovaci\'on MCIN/AEI/10.13039/501100011033 and the ERDF “A way of making Europe” through projects PID2021-125627OB-C32 and PID2024-158486OB-C32. We thank the anonymous referee for careful reading of this manuscript and useful suggestions.

\section*{Data Availability}
The \textit{TESS} photometric data and the high-resolution speckle imaging observations used in this work are publicly available through the Mikulski Archive for Space Telescopes (MAST; \url{https://mast.stsci.edu/}) and the ExoFOP--\textit{TESS} portal (\url{https://exofop.ipac.caltech.edu/tess/target.php?id=156514476}), respectively.  
The ground-based photometric follow-up data are also available on the ExoFOP--\textit{TESS} page.  
The radial velocity measurements underlying this article are provided in Table~\ref{tab:RV_measurements}.  
All other data supporting the findings of this study will be shared by the corresponding author upon reasonable request.






\bibliographystyle{mnras}
\bibliography{references}

\begin{thebibliography}{}
\makeatletter
\relax
\def\mn@urlcharsother{\let\do\@makeother \do\$\do\&\do\#\do\^\do\_\do\%\do\~}
\def\mn@doi{\begingroup\mn@urlcharsother \@ifnextchar [ {\mn@doi@} {\mn@doi@[]}}
\def\mn@doi@[#1]#2{\def\@tempa{#1}\ifx\@tempa\@empty \href {http://dx.doi.org/#2} {doi:#2}\else \href {http://dx.doi.org/#2} {#1}\fi \endgroup}
\def\mn@eprint#1#2{\mn@eprint@#1:#2::\@nil}
\def\mn@eprint@arXiv#1{\href {http://arxiv.org/abs/#1} {{\tt arXiv:#1}}}
\def\mn@eprint@dblp#1{\href {http://dblp.uni-trier.de/rec/bibtex/#1.xml} {dblp:#1}}
\def\mn@eprint@#1:#2:#3:#4\@nil{\def\@tempa {#1}\def\@tempb {#2}\def\@tempc {#3}\ifx \@tempc \@empty \let \@tempc \@tempb \let \@tempb \@tempa \fi \ifx \@tempb \@empty \def\@tempb {arXiv}\fi \@ifundefined {mn@eprint@\@tempb}{\@tempb:\@tempc}{\expandafter \expandafter \csname mn@eprint@\@tempb\endcsname \expandafter{\@tempc}}}

\bibitem[\protect\citeauthoryear{{Adams} \& {Laughlin}}{{Adams} \& {Laughlin}}{2006}]{Adams2006}
{Adams} F.~C.,  {Laughlin} G.,  2006, \mn@doi [\apj] {10.1086/506145}, \href {https://ui.adsabs.harvard.edu/abs/2006ApJ...649.1004A} {649, 1004}

\bibitem[\protect\citeauthoryear{Agol, Luger  \& Foreman-Mackey}{Agol et~al.}{2020}]{Agol_2020}
Agol E.,  Luger R.,   Foreman-Mackey D.,  2020, \mn@doi [The Astronomical Journal] {10.3847/1538-3881/ab4fee}, 159, 123

\bibitem[\protect\citeauthoryear{{Albrecht} et~al.,}{{Albrecht} et~al.}{2012}]{Albrecht2012}
{Albrecht} S.,  et~al., 2012, \mn@doi [\apj] {10.1088/0004-637X/757/1/18}, \href {https://ui.adsabs.harvard.edu/abs/2012ApJ...757...18A} {757, 18}

\bibitem[\protect\citeauthoryear{{Alibert}, {Mordasini}, {Benz}  \& {Winisdoerffer}}{{Alibert} et~al.}{2005}]{Alibert2005}
{Alibert} Y.,  {Mordasini} C.,  {Benz} W.,   {Winisdoerffer} C.,  2005, \mn@doi [\aap] {10.1051/0004-6361:20042032}, \href {https://ui.adsabs.harvard.edu/abs/2005A&A...434..343A} {434, 343}

\bibitem[\protect\citeauthoryear{{Aller}, {Lillo-Box}, {Jones}, {Miranda}  \& {Barcel{\'o} Forteza}}{{Aller} et~al.}{2020}]{tpfplot}
{Aller} A.,  {Lillo-Box} J.,  {Jones} D.,  {Miranda} L.~F.,   {Barcel{\'o} Forteza} S.,  2020, \mn@doi [\aap] {10.1051/0004-6361/201937118}, \href {https://ui.adsabs.harvard.edu/abs/2020A&A...635A.128A} {635, A128}

\bibitem[\protect\citeauthoryear{Baliwal et~al.,}{Baliwal et~al.}{2024}]{Baliwal2024}
Baliwal S.,  et~al., 2024, \mn@doi [A\&A] {10.1051/0004-6361/202450934}, 691, A12

\bibitem[\protect\citeauthoryear{{Baliwal} et~al.,}{{Baliwal} et~al.}{2025}]{toi6038}
{Baliwal} S.,  et~al., 2025, \mn@doi [\aj] {10.3847/1538-3881/ada959}, \href {https://ui.adsabs.harvard.edu/abs/2025AJ....169..147B} {169, 147}

\bibitem[\protect\citeauthoryear{{Baraffe}, {Chabrier}, {Allard}  \& {Hauschildt}}{{Baraffe} et~al.}{2002}]{Baraffe2002}
{Baraffe} I.,  {Chabrier} G.,  {Allard} F.,   {Hauschildt} P.~H.,  2002, \mn@doi [\aap] {10.1051/0004-6361:20011638}, \href {https://ui.adsabs.harvard.edu/abs/2002A&A...382..563B} {382, 563}

\bibitem[\protect\citeauthoryear{{Baraffe}, {Homeier}, {Allard}  \& {Chabrier}}{{Baraffe} et~al.}{2015}]{Baraffe2015}
{Baraffe} I.,  {Homeier} D.,  {Allard} F.,   {Chabrier} G.,  2015, \mn@doi [\aap] {10.1051/0004-6361/201425481}, \href {https://ui.adsabs.harvard.edu/abs/2015A&A...577A..42B} {577, A42}

\bibitem[\protect\citeauthoryear{{Baranne} et~al.,}{{Baranne} et~al.}{1996}]{Baranne1996}
{Baranne} A.,  et~al., 1996, \aaps, \href {https://ui.adsabs.harvard.edu/abs/1996A&AS..119..373B} {119, 373}

\bibitem[\protect\citeauthoryear{Barragán, Aigrain, Rajpaul  \& Zicher}{Barragán et~al.}{2021}]{citlalicue}
Barragán O.,  Aigrain S.,  Rajpaul V.~M.,   Zicher N.,  2021, \mn@doi [Monthly Notices of the Royal Astronomical Society] {10.1093/mnras/stab2889}, 509, 866

\bibitem[\protect\citeauthoryear{{Beatty}, {Morley}, {Curtis}, {Burrows}, {Davenport}  \& {Montet}}{{Beatty} et~al.}{2018}]{Beatty2018}
{Beatty} T.~G.,  {Morley} C.~V.,  {Curtis} J.~L.,  {Burrows} A.,  {Davenport} J. R.~A.,   {Montet} B.~T.,  2018, \mn@doi [\aj] {10.3847/1538-3881/aad697}, \href {https://ui.adsabs.harvard.edu/abs/2018AJ....156..168B} {156, 168}

\bibitem[\protect\citeauthoryear{{Bensby}, {Feltzing}  \& {Oey}}{{Bensby} et~al.}{2014}]{Bensby2014}
{Bensby} T.,  {Feltzing} S.,   {Oey} M.~S.,  2014, \mn@doi [\aap] {10.1051/0004-6361/201322631}, \href {https://ui.adsabs.harvard.edu/abs/2014A&A...562A..71B} {562, A71}

\bibitem[\protect\citeauthoryear{{Bouchy} et~al.,}{{Bouchy} et~al.}{2011}]{Bouchy2011}
{Bouchy} F.,  et~al., 2011, \mn@doi [\aap] {10.1051/0004-6361/201015276}, \href {https://ui.adsabs.harvard.edu/abs/2011A&A...525A..68B} {525, A68}

\bibitem[\protect\citeauthoryear{{Brown} et~al.,}{{Brown} et~al.}{2013}]{Brown2013}
{Brown} T.~M.,  et~al., 2013, \mn@doi [\pasp] {10.1086/673168}, \href {https://ui.adsabs.harvard.edu/abs/2013PASP..125.1031B} {125, 1031}

\bibitem[\protect\citeauthoryear{{Buchhave} et~al.,}{{Buchhave} et~al.}{2010}]{Buchhave2010}
{Buchhave} L.~A.,  et~al., 2010, \mn@doi [\apj] {10.1088/0004-637X/720/2/1118}, \href {https://ui.adsabs.harvard.edu/abs/2010ApJ...720.1118B} {720, 1118}

\bibitem[\protect\citeauthoryear{Buchhave et~al.,}{Buchhave et~al.}{2012}]{Buchhave2012}
Buchhave L.~A.,  et~al., 2012, Nature, 486, 375

\bibitem[\protect\citeauthoryear{{Burrows}, {Hubeny}, {Budaj}  \& {Hubbard}}{{Burrows} et~al.}{2007}]{Burrows2007}
{Burrows} A.,  {Hubeny} I.,  {Budaj} J.,   {Hubbard} W.~B.,  2007, \mn@doi [\apj] {10.1086/514326}, \href {https://ui.adsabs.harvard.edu/abs/2007ApJ...661..502B} {661, 502}

\bibitem[\protect\citeauthoryear{{Burrows}, {Heng}  \& {Nampaisarn}}{{Burrows} et~al.}{2011}]{Burrows2011}
{Burrows} A.,  {Heng} K.,   {Nampaisarn} T.,  2011, \mn@doi [\apj] {10.1088/0004-637X/736/1/47}, \href {https://ui.adsabs.harvard.edu/abs/2011ApJ...736...47B} {736, 47}

\bibitem[\protect\citeauthoryear{{Butters} et~al.,}{{Butters} et~al.}{2010}]{Butters2010}
{Butters} O.~W.,  et~al., 2010, \mn@doi [\aap] {10.1051/0004-6361/201015655}, \href {https://ui.adsabs.harvard.edu/abs/2010A&A...520L..10B} {520, L10}

\bibitem[\protect\citeauthoryear{{Castro-Gonz{\'a}lez} et~al.,}{{Castro-Gonz{\'a}lez} et~al.}{2024}]{CastroGonzalez2024b}
{Castro-Gonz{\'a}lez} A.,  et~al., 2024, \mn@doi [\aap] {10.1051/0004-6361/202451656}, \href {https://ui.adsabs.harvard.edu/abs/2024A&A...691A.233C} {691, A233}

\bibitem[\protect\citeauthoryear{{Chabrier}, {Baraffe}, {Allard}  \& {Hauschildt}}{{Chabrier} et~al.}{2000}]{Chabrier2000}
{Chabrier} G.,  {Baraffe} I.,  {Allard} F.,   {Hauschildt} P.,  2000, \mn@doi [\apj] {10.1086/309513}, \href {https://ui.adsabs.harvard.edu/abs/2000ApJ...542..464C} {542, 464}

\bibitem[\protect\citeauthoryear{{Chabrier}, {Baraffe}, {Phillips}  \& {Debras}}{{Chabrier} et~al.}{2023}]{Chabrier2023}
{Chabrier} G.,  {Baraffe} I.,  {Phillips} M.,   {Debras} F.,  2023, \mn@doi [\aap] {10.1051/0004-6361/202243832}, \href {https://ui.adsabs.harvard.edu/abs/2023A&A...671A.119C} {671, A119}

\bibitem[\protect\citeauthoryear{{Chakraborty}, {Thapa}, {Kumar}, {Neelam}, {Sharma}  \& {Roy}}{{Chakraborty} et~al.}{2018}]{Chakraborty2018}
{Chakraborty} A.,  {Thapa} N.,  {Kumar} K.,  {Neelam} P. J.~S.~S.~V.,  {Sharma} R.,   {Roy} A.,  2018, in {Evans} C.~J.,  {Simard} L.,   {Takami} H.,  eds,  Society of Photo-Optical Instrumentation Engineers (SPIE) Conference Series Vol. 10702, Ground-based and Airborne Instrumentation for Astronomy VII. p. 107026G, \mn@doi{10.1117/12.2313055}

\bibitem[\protect\citeauthoryear{Chakraborty et~al.,}{Chakraborty et~al.}{2024}]{Chakraborty2024}
Chakraborty A.,  et~al., 2024, \mn@doi [Bulletin de la Société Royale des Sciences de Liège] {10.25518/0037-9565.11602}, p. 68–88

\bibitem[\protect\citeauthoryear{Chaturvedi, Chakraborty, Anandarao, Roy  \& Mahadevan}{Chaturvedi et~al.}{2016}]{Chaturvedi2016}
Chaturvedi P.,  Chakraborty A.,  Anandarao B.~G.,  Roy A.,   Mahadevan S.,  2016, \mn@doi [Monthly Notices of the Royal Astronomical Society] {10.1093/mnras/stw1560}, 462, 554

\bibitem[\protect\citeauthoryear{{Choi}, {Dotter}, {Conroy}, {Cantiello}, {Paxton}  \& {Johnson}}{{Choi} et~al.}{2016}]{mist_choi}
{Choi} J.,  {Dotter} A.,  {Conroy} C.,  {Cantiello} M.,  {Paxton} B.,   {Johnson} B.~D.,  2016, \mn@doi [\apj] {10.3847/0004-637X/823/2/102}, \href {https://ui.adsabs.harvard.edu/abs/2016ApJ...823..102C} {823, 102}

\bibitem[\protect\citeauthoryear{{Ciardi}, {Beichman}, {Horch}  \& {Howell}}{{Ciardi} et~al.}{2015}]{Ciardi2015}
{Ciardi} D.~R.,  {Beichman} C.~A.,  {Horch} E.~P.,   {Howell} S.~B.,  2015, \mn@doi [\apj] {10.1088/0004-637X/805/1/16}, \href {https://ui.adsabs.harvard.edu/abs/2015ApJ...805...16C} {805, 16}

\bibitem[\protect\citeauthoryear{{Claret}}{{Claret}}{2017}]{Claret_tess}
{Claret} A.,  2017, \mn@doi [\aap] {10.1051/0004-6361/201629705}, \href {https://ui.adsabs.harvard.edu/abs/2017A&A...600A..30C} {600, A30}

\bibitem[\protect\citeauthoryear{{Claret} \& {Bloemen}}{{Claret} \& {Bloemen}}{2011}]{Claret}
{Claret} A.,  {Bloemen} S.,  2011, \mn@doi [\aap] {10.1051/0004-6361/201116451}, \href {https://ui.adsabs.harvard.edu/abs/2011A&A...529A..75C} {529, A75}

\bibitem[\protect\citeauthoryear{{Collins}, {Kielkopf}, {Stassun}  \& {Hessman}}{{Collins} et~al.}{2017}]{Collins2017}
{Collins} K.~A.,  {Kielkopf} J.~F.,  {Stassun} K.~G.,   {Hessman} F.~V.,  2017, \mn@doi [\aj] {10.3847/1538-3881/153/2/77}, \href {http://adsabs.harvard.edu/abs/2017AJ....153...77C} {153, 77}

\bibitem[\protect\citeauthoryear{{Cutri} et~al.,}{{Cutri} et~al.}{2003}]{JHK}
{Cutri} R.~M.,  et~al., 2003, VizieR Online Data Catalog, \href {https://ui.adsabs.harvard.edu/abs/2003yCat.2246....0C} {p. II/246}

\bibitem[\protect\citeauthoryear{{Cutri} et~al.,}{{Cutri} et~al.}{2021}]{ALLWISE}
{Cutri} R.~M.,  et~al., 2021, VizieR Online Data Catalog, \href {https://ui.adsabs.harvard.edu/abs/2014yCat.2328....0C} {p. II/328}

\bibitem[\protect\citeauthoryear{{Dekany} et~al.,}{{Dekany} et~al.}{2013}]{dekany2013}
{Dekany} R.,  et~al., 2013, \mn@doi [\apj] {10.1088/0004-637X/776/2/130}, \href {https://ui.adsabs.harvard.edu/abs/2013ApJ...776..130D} {776, 130}

\bibitem[\protect\citeauthoryear{{Dotter}}{{Dotter}}{2016}]{mist_dotter}
{Dotter} A.,  2016, \mn@doi [\apjs] {10.3847/0067-0049/222/1/8}, \href {https://ui.adsabs.harvard.edu/abs/2016ApJS..222....8D} {222, 8}

\bibitem[\protect\citeauthoryear{{Eastman} et~al.,}{{Eastman} et~al.}{2019}]{exofast}
{Eastman} J.~D.,  et~al., 2019, arXiv e-prints, \href {https://ui.adsabs.harvard.edu/abs/2019arXiv190709480E} {p. arXiv:1907.09480}

\bibitem[\protect\citeauthoryear{F\H{u}r\'esz}{F\H{u}r\'esz}{2008}]{gaborthesis}
F\H{u}r\'esz G.,  2008, PhD thesis, University of Szeged, Hungary

\bibitem[\protect\citeauthoryear{{Ford}}{{Ford}}{2006}]{Gelman_rubin2006}
{Ford} E.~B.,  2006, \mn@doi [\apj] {10.1086/500802}, \href {https://ui.adsabs.harvard.edu/abs/2006ApJ...642..505F} {642, 505}

\bibitem[\protect\citeauthoryear{{Furlan} \& {Howell}}{{Furlan} \& {Howell}}{2017}]{Furlan2017}
{Furlan} E.,  {Howell} S.~B.,  2017, \mn@doi [\aj] {10.3847/1538-3881/aa7b70}, \href {https://ui.adsabs.harvard.edu/abs/2017AJ....154...66F} {154, 66}

\bibitem[\protect\citeauthoryear{{Furlan} \& {Howell}}{{Furlan} \& {Howell}}{2020}]{Furlan2020}
{Furlan} E.,  {Howell} S.~B.,  2020, \mn@doi [\apj] {10.3847/1538-4357/ab9c9c}, \href {https://ui.adsabs.harvard.edu/abs/2020ApJ...898...47F} {898, 47}

\bibitem[\protect\citeauthoryear{{Furlan} et~al.,}{{Furlan} et~al.}{2017}]{Furlan2017_ciardi}
{Furlan} E.,  et~al., 2017, \mn@doi [\aj] {10.3847/1538-3881/153/2/71}, \href {https://ui.adsabs.harvard.edu/abs/2017AJ....153...71F} {153, 71}

\bibitem[\protect\citeauthoryear{{Gagn{\'e}} et~al.,}{{Gagn{\'e}} et~al.}{2018}]{Gagne2018}
{Gagn{\'e}} J.,  et~al., 2018, \mn@doi [\apj] {10.3847/1538-4357/aaae09}, \href {https://ui.adsabs.harvard.edu/abs/2018ApJ...856...23G} {856, 23}

\bibitem[\protect\citeauthoryear{{Gaia Collaboration} et~al.,}{{Gaia Collaboration} et~al.}{2023}]{gaia2023}
{Gaia Collaboration} et~al., 2023, \mn@doi [\aap] {10.1051/0004-6361/202243940}, \href {https://ui.adsabs.harvard.edu/abs/2023A&A...674A...1G} {674, A1}

\bibitem[\protect\citeauthoryear{{Gardner} et~al.,}{{Gardner} et~al.}{2006}]{Gardner_2006}
{Gardner} J.~P.,  et~al., 2006, \mn@doi [\ssr] {10.1007/s11214-006-8315-7}, \href {https://ui.adsabs.harvard.edu/abs/2006SSRv..123..485G} {123, 485}

\bibitem[\protect\citeauthoryear{{Gelman} \& {Rubin}}{{Gelman} \& {Rubin}}{1992}]{Gelman1992}
{Gelman} A.,  {Rubin} D.~B.,  1992, \mn@doi [Statistical Science] {10.1214/ss/1177011136}, \href {https://ui.adsabs.harvard.edu/abs/1992StaSc...7..457G} {7, 457}

\bibitem[\protect\citeauthoryear{{Grieves} et~al.,}{{Grieves} et~al.}{2017}]{Grieves2017}
{Grieves} N.,  et~al., 2017, \mn@doi [\mnras] {10.1093/mnras/stx334}, \href {https://ui.adsabs.harvard.edu/abs/2017MNRAS.467.4264G} {467, 4264}

\bibitem[\protect\citeauthoryear{{Grieves} et~al.,}{{Grieves} et~al.}{2021}]{Grieves2021}
{Grieves} N.,  et~al., 2021, \mn@doi [\aap] {10.1051/0004-6361/202141145}, \href {https://ui.adsabs.harvard.edu/abs/2021A&A...652A.127G} {652, A127}

\bibitem[\protect\citeauthoryear{{Grunblatt} et~al.,}{{Grunblatt} et~al.}{2016}]{Grunblatt2016}
{Grunblatt} S.~K.,  et~al., 2016, \mn@doi [\aj] {10.3847/0004-6256/152/6/185}, \href {https://ui.adsabs.harvard.edu/abs/2016AJ....152..185G} {152, 185}

\bibitem[\protect\citeauthoryear{{Grunblatt} et~al.,}{{Grunblatt} et~al.}{2018}]{Grunblatt2018}
{Grunblatt} S.~K.,  et~al., 2018, \mn@doi [\apjl] {10.3847/2041-8213/aacc67}, \href {https://ui.adsabs.harvard.edu/abs/2018ApJ...861L...5G} {861, L5}

\bibitem[\protect\citeauthoryear{{Guerrero} et~al.,}{{Guerrero} et~al.}{2021}]{Guerrero2021}
{Guerrero} N.~M.,  et~al., 2021, \mn@doi [\apjs] {10.3847/1538-4365/abefe1}, \href {https://ui.adsabs.harvard.edu/abs/2021ApJS..254...39G} {254, 39}

\bibitem[\protect\citeauthoryear{{Halbwachs}, {Mayor}, {Udry}  \& {Arenou}}{{Halbwachs} et~al.}{2003}]{Halbwachs2003}
{Halbwachs} J.~L.,  {Mayor} M.,  {Udry} S.,   {Arenou} F.,  2003, \mn@doi [\aap] {10.1051/0004-6361:20021507}, \href {https://ui.adsabs.harvard.edu/abs/2003A&A...397..159H} {397, 159}

\bibitem[\protect\citeauthoryear{{Hayward}, {Brandl}, {Pirger}, {Blacken}, {Gull}, {Schoenwald}  \& {Houck}}{{Hayward} et~al.}{2001}]{hayward2001}
{Hayward} T.~L.,  {Brandl} B.,  {Pirger} B.,  {Blacken} C.,  {Gull} G.~E.,  {Schoenwald} J.,   {Houck} J.~R.,  2001, \mn@doi [\pasp] {10.1086/317969}, \href {https://ui.adsabs.harvard.edu/abs/2001PASP..113..105H} {113, 105}

\bibitem[\protect\citeauthoryear{{Henderson} et~al.,}{{Henderson} et~al.}{2024}]{Henderson2024}
{Henderson} B.~A.,  et~al., 2024, \mn@doi [\mnras] {10.1093/mnras/stae1940}, \href {https://ui.adsabs.harvard.edu/abs/2024MNRAS.533.2823H} {533, 2823}

\bibitem[\protect\citeauthoryear{{Hennebelle} \& {Chabrier}}{{Hennebelle} \& {Chabrier}}{2008}]{Hennebelle2008}
{Hennebelle} P.,  {Chabrier} G.,  2008, \mn@doi [\apj] {10.1086/589916}, \href {https://ui.adsabs.harvard.edu/abs/2008ApJ...684..395H} {684, 395}

\bibitem[\protect\citeauthoryear{{H{\o}g} et~al.,}{{H{\o}g} et~al.}{2000}]{tycho}
{H{\o}g} E.,  et~al., 2000, \aap, \href {https://ui.adsabs.harvard.edu/abs/2000A&A...355L..27H} {355, L27}

\bibitem[\protect\citeauthoryear{{Howell}, {Everett}, {Sherry}, {Horch}  \& {Ciardi}}{{Howell} et~al.}{2011}]{Howell2011}
{Howell} S.~B.,  {Everett} M.~E.,  {Sherry} W.,  {Horch} E.,   {Ciardi} D.~R.,  2011, \mn@doi [\aj] {10.1088/0004-6256/142/1/19}, \href {https://ui.adsabs.harvard.edu/abs/2011AJ....142...19H} {142, 19}

\bibitem[\protect\citeauthoryear{{Howell}, {Scott}, {Matson}, {Everett}, {Furlan}, {Gnilka}, {Ciardi}  \& {Lester}}{{Howell} et~al.}{2021}]{Howell2021}
{Howell} S.~B.,  {Scott} N.~J.,  {Matson} R.~A.,  {Everett} M.~E.,  {Furlan} E.,  {Gnilka} C.~L.,  {Ciardi} D.~R.,   {Lester} K.~V.,  2021, \mn@doi [Frontiers in Astronomy and Space Sciences] {10.3389/fspas.2021.635864}, \href {https://ui.adsabs.harvard.edu/abs/2021FrASS...8...10H} {8, 10}

\bibitem[\protect\citeauthoryear{{Hut}}{{Hut}}{1981}]{Hut1981}
{Hut} P.,  1981, \aap, \href {https://ui.adsabs.harvard.edu/abs/1981A&A....99..126H} {99, 126}

\bibitem[\protect\citeauthoryear{{Jackson}, {Greenberg}  \& {Barnes}}{{Jackson} et~al.}{2008}]{Jackson+2008}
{Jackson} B.,  {Greenberg} R.,   {Barnes} R.,  2008, \mn@doi [\apj] {10.1086/529187}, \href {https://ui.adsabs.harvard.edu/abs/2008ApJ...678.1396J} {678, 1396}

\bibitem[\protect\citeauthoryear{{Jackson}, {Barnes}  \& {Greenberg}}{{Jackson} et~al.}{2009}]{Jackson+2009}
{Jackson} B.,  {Barnes} R.,   {Greenberg} R.,  2009, \mn@doi [\apj] {10.1088/0004-637X/698/2/1357}, \href {https://ui.adsabs.harvard.edu/abs/2009ApJ...698.1357J} {698, 1357}

\bibitem[\protect\citeauthoryear{Jenkins et~al.,}{Jenkins et~al.}{2016}]{spoc}
Jenkins J.,  et~al., 2016, in Software and Cyber infrastructure for Astronomy IV. , \mn@doi{10.1117/12.2233418}

\bibitem[\protect\citeauthoryear{{Jensen}}{{Jensen}}{2013}]{Jensen2013}
{Jensen} E.,  2013, {Tapir: A web interface for transit/eclipse observability}, Astrophysics Source Code Library (\mn@eprint {ascl} {1306.007})

\bibitem[\protect\citeauthoryear{{Kempton} et~al.,}{{Kempton} et~al.}{2018}]{Kempton_2018}
{Kempton} E. M.-R.,  et~al., 2018, \mn@doi [\pasp] {10.1088/1538-3873/aadf6f}, \href {https://ui.adsabs.harvard.edu/abs/2018PASP..130k4401K} {130, 114401}

\bibitem[\protect\citeauthoryear{{Khandelwal} et~al.,}{{Khandelwal} et~al.}{2022}]{toi1789}
{Khandelwal} A.,  et~al., 2022, \mn@doi [\mnras] {10.1093/mnras/stab2970}, \href {https://ui.adsabs.harvard.edu/abs/2022MNRAS.509.3339K} {509, 3339}

\bibitem[\protect\citeauthoryear{{Kiefer}, {H{\'e}brard}, {Lecavelier des Etangs}, {Martioli}, {Dalal}  \& {Vidal-Madjar}}{{Kiefer} et~al.}{2021}]{Kiefer2021}
{Kiefer} F.,  {H{\'e}brard} G.,  {Lecavelier des Etangs} A.,  {Martioli} E.,  {Dalal} S.,   {Vidal-Madjar} A.,  2021, \mn@doi [\aap] {10.1051/0004-6361/202039168}, \href {https://ui.adsabs.harvard.edu/abs/2021A&A...645A...7K} {645, A7}

\bibitem[\protect\citeauthoryear{{Kraft}}{{Kraft}}{1967}]{Kraft1967}
{Kraft} R.~P.,  1967, \mn@doi [\apj] {10.1086/149359}, \href {https://ui.adsabs.harvard.edu/abs/1967ApJ...150..551K} {150, 551}

\bibitem[\protect\citeauthoryear{{Kratter} \& {Lodato}}{{Kratter} \& {Lodato}}{2016}]{Kratter2016}
{Kratter} K.,  {Lodato} G.,  2016, \mn@doi [\araa] {10.1146/annurev-astro-081915-023307}, \href {https://ui.adsabs.harvard.edu/abs/2016ARA&A..54..271K} {54, 271}

\bibitem[\protect\citeauthoryear{Kreidberg}{Kreidberg}{2015}]{batman}
Kreidberg L.,  2015, \mn@doi [Publications of the Astronomical Society of the Pacific] {10.1086/683602}, 127, 1161–1165

\bibitem[\protect\citeauthoryear{{Kurucz}}{{Kurucz}}{1992}]{Kurucz1992}
{Kurucz} R.~L.,  1992, in {Barbuy} B.,  {Renzini} A.,  eds,  IAU Symposium Vol. 149, The Stellar Populations of Galaxies. p.~225

\bibitem[\protect\citeauthoryear{{Lanza}, {Damiani}  \& {Gandolfi}}{{Lanza} et~al.}{2011}]{Lanza+2011}
{Lanza} A.~F.,  {Damiani} C.,   {Gandolfi} D.,  2011, \mn@doi [\aap] {10.1051/0004-6361/201016144}, \href {https://ui.adsabs.harvard.edu/abs/2011A&A...529A..50L} {529, A50}

\bibitem[\protect\citeauthoryear{{Leggett}}{{Leggett}}{1992}]{Leggett1992}
{Leggett} S.~K.,  1992, \mn@doi [\apjs] {10.1086/191720}, \href {https://ui.adsabs.harvard.edu/abs/1992ApJS...82..351L} {82, 351}

\bibitem[\protect\citeauthoryear{{Lester} et~al.,}{{Lester} et~al.}{2021}]{Lester2021}
{Lester} K.~V.,  et~al., 2021, \mn@doi [\aj] {10.3847/1538-3881/ac0d06}, \href {https://ui.adsabs.harvard.edu/abs/2021AJ....162...75L} {162, 75}

\bibitem[\protect\citeauthoryear{{Lillo-Box}, {Barrado}  \& {Bouy}}{{Lillo-Box} et~al.}{2014}]{LilloBox2014}
{Lillo-Box} J.,  {Barrado} D.,   {Bouy} H.,  2014, \mn@doi [\aap] {10.1051/0004-6361/201423497}, \href {https://ui.adsabs.harvard.edu/abs/2014A&A...566A.103L} {566, A103}

\bibitem[\protect\citeauthoryear{{Lin} et~al.,}{{Lin} et~al.}{2023}]{Lin2023}
{Lin} Z.,  et~al., 2023, \mn@doi [\mnras] {10.1093/mnras/stad1745}, \href {https://ui.adsabs.harvard.edu/abs/2023MNRAS.523.6162L} {523, 6162}

\bibitem[\protect\citeauthoryear{{Lindegren} et~al.,}{{Lindegren} et~al.}{2021}]{Lindegren_2021}
{Lindegren} et~al., 2021, \mn@doi [A\&A] {10.1051/0004-6361/202039709}, 649, A2

\bibitem[\protect\citeauthoryear{{Lucy} \& {Sweeney}}{{Lucy} \& {Sweeney}}{1971}]{Lucy1971}
{Lucy} L.~B.,  {Sweeney} M.~A.,  1971, \mn@doi [\aj] {10.1086/111159}, \href {https://ui.adsabs.harvard.edu/abs/1971AJ.....76..544L} {76, 544}

\bibitem[\protect\citeauthoryear{{Ma} \& {Ge}}{{Ma} \& {Ge}}{2014}]{Ma2014}
{Ma} B.,  {Ge} J.,  2014, \mn@doi [\mnras] {10.1093/mnras/stu134}, \href {https://ui.adsabs.harvard.edu/abs/2014MNRAS.439.2781M} {439, 2781}

\bibitem[\protect\citeauthoryear{{Mandel} \& {Agol}}{{Mandel} \& {Agol}}{2002}]{Mandel2002}
{Mandel} K.,  {Agol} E.,  2002, \mn@doi [\apjl] {10.1086/345520}, \href {https://ui.adsabs.harvard.edu/abs/2002ApJ...580L.171M} {580, L171}

\bibitem[\protect\citeauthoryear{{Masuda} \& {Winn}}{{Masuda} \& {Winn}}{2020}]{Masuda2020}
{Masuda} K.,  {Winn} J.~N.,  2020, \mn@doi [\aj] {10.3847/1538-3881/ab65be}, \href {https://ui.adsabs.harvard.edu/abs/2020AJ....159...81M} {159, 81}

\bibitem[\protect\citeauthoryear{{Matson}, {Howell}, {Horch}  \& {Everett}}{{Matson} et~al.}{2018}]{Matson2018}
{Matson} R.,  {Howell} S.,  {Horch} E.,   {Everett} M.,  2018, in American Astronomical Society Meeting Abstracts \#231. p. 109.02

\bibitem[\protect\citeauthoryear{{McCully}, {Volgenau}, {Harbeck}, {Lister}, {Saunders}, {Turner}, {Siiverd}  \& {Bowman}}{{McCully} et~al.}{2018}]{McCully2018}
{McCully} C.,  {Volgenau} N.~H.,  {Harbeck} D.-R.,  {Lister} T.~A.,  {Saunders} E.~S.,  {Turner} M.~L.,  {Siiverd} R.~J.,   {Bowman} M.,  2018, in \procspie. p. 107070K (\mn@eprint {arXiv} {1811.04163}), \mn@doi{10.1117/12.2314340}

\bibitem[\protect\citeauthoryear{{Mordasini}, {Alibert}, {Benz}  \& {Naef}}{{Mordasini} et~al.}{2009}]{Mordasini2009}
{Mordasini} C.,  {Alibert} Y.,  {Benz} W.,   {Naef} D.,  2009, \mn@doi [\aap] {10.1051/0004-6361/200810697}, \href {https://ui.adsabs.harvard.edu/abs/2009A&A...501.1161M} {501, 1161}

\bibitem[\protect\citeauthoryear{{Morley} et~al.,}{{Morley} et~al.}{2024}]{Morley2024}
{Morley} C.~V.,  et~al., 2024, \mn@doi [\apj] {10.3847/1538-4357/ad71d5}, \href {https://ui.adsabs.harvard.edu/abs/2024ApJ...975...59M} {975, 59}

\bibitem[\protect\citeauthoryear{{Mukherjee}, {Fortney}, {Carmichael}, {Davis}  \& {Thorngren}}{{Mukherjee} et~al.}{2025}]{Mukherjee2025}
{Mukherjee} S.,  {Fortney} J.~J.,  {Carmichael} T.~W.,  {Davis} C.~E.,   {Thorngren} D.~P.,  2025, \mn@doi [arXiv e-prints] {10.48550/arXiv.2512.08249}, \href {https://ui.adsabs.harvard.edu/abs/2025arXiv251208249M} {p. arXiv:2512.08249}

\bibitem[\protect\citeauthoryear{{Narita} et~al.,}{{Narita} et~al.}{2019}]{Narita2019}
{Narita} N.,  et~al., 2019, \mn@doi [Journal of Astronomical Telescopes, Instruments, and Systems] {10.1117/1.JATIS.5.1.015001}, \href {https://ui.adsabs.harvard.edu/abs/2019JATIS...5a5001N} {5, 015001}

\bibitem[\protect\citeauthoryear{{Narita} et~al.,}{{Narita} et~al.}{2020}]{Narita:2020}
{Narita} N.,  et~al., 2020, in Society of Photo-Optical Instrumentation Engineers (SPIE) Conference Series. p. 114475K, \mn@doi{10.1117/12.2559947}

\bibitem[\protect\citeauthoryear{{Nguyen}, {Caldwell}, {Twicken}, {Striegel}, {Ting}, {Williams}  \& {Jenkins}}{{Nguyen} et~al.}{2022}]{Nguyen2022}
{Nguyen} K.~T.,  {Caldwell} D.~A.,  {Twicken} J.~D.,  {Striegel} S.~L.,  {Ting} E.~B.,  {Williams} R.~H.,   {Jenkins} J.~M.,  2022, \mn@doi [Research Notes of the American Astronomical Society] {10.3847/2515-5172/ac983a}, \href {https://ui.adsabs.harvard.edu/abs/2022RNAAS...6..207N} {6, 207}

\bibitem[\protect\citeauthoryear{{Padoan} \& {Nordlund}}{{Padoan} \& {Nordlund}}{2004}]{Padoan2004}
{Padoan} P.,  {Nordlund} {\r{A}}.,  2004, \mn@doi [\apj] {10.1086/345413}, \href {https://ui.adsabs.harvard.edu/abs/2004ApJ...617..559P} {617, 559}

\bibitem[\protect\citeauthoryear{{Page} et~al.,}{{Page} et~al.}{2024}]{Page2024}
{Page} E.,  et~al., 2024, \mn@doi [\aj] {10.3847/1538-3881/ad1a18}, \href {https://ui.adsabs.harvard.edu/abs/2024AJ....167..109P} {167, 109}

\bibitem[\protect\citeauthoryear{{Parviainen} et~al.,}{{Parviainen} et~al.}{2019}]{Parviainen2019}
{Parviainen} H.,  et~al., 2019, \mn@doi [\aap] {10.1051/0004-6361/201935709}, \href {https://ui.adsabs.harvard.edu/abs/2019A&A...630A..89P} {630, A89}

\bibitem[\protect\citeauthoryear{{Phillips} et~al.,}{{Phillips} et~al.}{2020}]{Phillips2020}
{Phillips} M.~W.,  et~al., 2020, \mn@doi [\aap] {10.1051/0004-6361/201937381}, \href {https://ui.adsabs.harvard.edu/abs/2020A&A...637A..38P} {637, A38}

\bibitem[\protect\citeauthoryear{{Psaridi} et~al.,}{{Psaridi} et~al.}{2022}]{Psaridi2022}
{Psaridi} A.,  et~al., 2022, \mn@doi [\aap] {10.1051/0004-6361/202243454}, \href {https://ui.adsabs.harvard.edu/abs/2022A&A...664A..94P} {664, A94}

\bibitem[\protect\citeauthoryear{{Rodriguez} et~al.,}{{Rodriguez} et~al.}{2023}]{Rodriguez2023}
{Rodriguez} J.~E.,  et~al., 2023, \mn@doi [\mnras] {10.1093/mnras/stad595}, \href {https://ui.adsabs.harvard.edu/abs/2023MNRAS.521.2765R} {521, 2765}

\bibitem[\protect\citeauthoryear{{Saunders} et~al.,}{{Saunders} et~al.}{2022}]{Saunders2022}
{Saunders} N.,  et~al., 2022, \mn@doi [\aj] {10.3847/1538-3881/ac38a1}, \href {https://ui.adsabs.harvard.edu/abs/2022AJ....163...53S} {163, 53}

\bibitem[\protect\citeauthoryear{{Schlafly} \& {Finkbeiner}}{{Schlafly} \& {Finkbeiner}}{2011}]{extinction}
{Schlafly} E.~F.,  {Finkbeiner} D.~P.,  2011, \mn@doi [\apj] {10.1088/0004-637X/737/2/103}, \href {https://ui.adsabs.harvard.edu/abs/2011ApJ...737..103S} {737, 103}

\bibitem[\protect\citeauthoryear{{Schlaufman}}{{Schlaufman}}{2018}]{Schlaufman2018}
{Schlaufman} K.~C.,  2018, \mn@doi [\apj] {10.3847/1538-4357/aa961c}, \href {https://ui.adsabs.harvard.edu/abs/2018ApJ...853...37S} {853, 37}

\bibitem[\protect\citeauthoryear{{Sch{\"o}nrich}, {Binney}  \& {Dehnen}}{{Sch{\"o}nrich} et~al.}{2010}]{Sch2010}
{Sch{\"o}nrich} R.,  {Binney} J.,   {Dehnen} W.,  2010, \mn@doi [\mnras] {10.1111/j.1365-2966.2010.16253.x}, \href {https://ui.adsabs.harvard.edu/abs/2010MNRAS.403.1829S} {403, 1829}

\bibitem[\protect\citeauthoryear{{Scott}, {Howell}, {Horch}  \& {Everett}}{{Scott} et~al.}{2018}]{scott2018}
{Scott} N.~J.,  {Howell} S.~B.,  {Horch} E.~P.,   {Everett} M.~E.,  2018, \mn@doi [\pasp] {10.1088/1538-3873/aab484}, \href {https://ui.adsabs.harvard.edu/abs/2018PASP..130e4502S} {130, 054502}

\bibitem[\protect\citeauthoryear{{Scott} et~al.,}{{Scott} et~al.}{2021}]{Scott2021}
{Scott} N.~J.,  et~al., 2021, \mn@doi [Frontiers in Astronomy and Space Sciences] {10.3389/fspas.2021.716560}, \href {https://ui.adsabs.harvard.edu/abs/2021FrASS...8..138S} {8, 138}

\bibitem[\protect\citeauthoryear{{Smith} et~al.,}{{Smith} et~al.}{2012}]{smith_2012}
{Smith} J.~C.,  et~al., 2012, \mn@doi [\pasp] {10.1086/667697}, \href {https://ui.adsabs.harvard.edu/abs/2012PASP..124.1000S} {124, 1000}

\bibitem[\protect\citeauthoryear{{Southworth}}{{Southworth}}{2011}]{tepcat}
{Southworth} J.,  2011, \mn@doi [\mnras] {10.1111/j.1365-2966.2011.19399.x}, \href {https://ui.adsabs.harvard.edu/abs/2011MNRAS.417.2166S} {417, 2166}

\bibitem[\protect\citeauthoryear{{Spiegel}, {Burrows}  \& {Milsom}}{{Spiegel} et~al.}{2011}]{Spiegel2011}
{Spiegel} D.~S.,  {Burrows} A.,   {Milsom} J.~A.,  2011, \mn@doi [\apj] {10.1088/0004-637X/727/1/57}, \href {https://ui.adsabs.harvard.edu/abs/2011ApJ...727...57S} {727, 57}

\bibitem[\protect\citeauthoryear{{Stassun} \& {Torres}}{{Stassun} \& {Torres}}{2016}]{SED1}
{Stassun} K.~G.,  {Torres} G.,  2016, \mn@doi [\apjl] {10.3847/2041-8205/831/1/L6}, \href {https://ui.adsabs.harvard.edu/abs/2016ApJ...831L...6S} {831, L6}

\bibitem[\protect\citeauthoryear{{Stassun} et~al.,}{{Stassun} et~al.}{2018}]{2018AJ....156..102S}
{Stassun} K.~G.,  et~al., 2018, \mn@doi [\aj] {10.3847/1538-3881/aad050}, \href {https://ui.adsabs.harvard.edu/abs/2018AJ....156..102S} {156, 102}

\bibitem[\protect\citeauthoryear{{Strakhov}, {Safonov}  \& {Cheryasov}}{{Strakhov} et~al.}{2023}]{Strakhov2023}
{Strakhov} I.~A.,  {Safonov} B.~S.,   {Cheryasov} D.~V.,  2023, \mn@doi [Astrophysical Bulletin] {10.1134/S1990341323020104}, \href {https://ui.adsabs.harvard.edu/abs/2023AstBu..78..234S} {78, 234}

\bibitem[\protect\citeauthoryear{Stumpe, Smith, Catanzarite, Cleve, Jenkins, Twicken  \& Girouard}{Stumpe et~al.}{2014}]{Stumpe_2014}
Stumpe M.~C.,  Smith J.~C.,  Catanzarite J.~H.,  Cleve J. E.~V.,  Jenkins J.~M.,  Twicken J.~D.,   Girouard F.~R.,  2014, \mn@doi [Publications of the Astronomical Society of the Pacific] {10.1086/674989}, 126, 100

\bibitem[\protect\citeauthoryear{{Tayar}, {Stassun}  \& {Corsaro}}{{Tayar} et~al.}{2019}]{Tayar2019}
{Tayar} J.,  {Stassun} K.~G.,   {Corsaro} E.,  2019, \mn@doi [\apj] {10.3847/1538-4357/ab3db1}, \href {https://ui.adsabs.harvard.edu/abs/2019ApJ...883..195T} {883, 195}

\bibitem[\protect\citeauthoryear{{Tinetti}, {Eccleston}, {Lueftinger}, {Salvignol}, {Fahmy}  \& {Alves de Oliveira}}{{Tinetti} et~al.}{2022}]{Tinetti_2018}
{Tinetti} G.,  {Eccleston} P.,  {Lueftinger} T.,  {Salvignol} J.-C.,  {Fahmy} S.,   {Alves de Oliveira} C.,  2022, in European Planetary Science Congress. pp EPSC2022--1114 (\mn@eprint {arXiv} {2104.04824}), \mn@doi{10.5194/epsc2022-1114}

\bibitem[\protect\citeauthoryear{{Torres}, {Winn}  \& {Holman}}{{Torres} et~al.}{2008}]{Torres2008}
{Torres} G.,  {Winn} J.~N.,   {Holman} M.~J.,  2008, \mn@doi [\apj] {10.1086/529429}, \href {https://ui.adsabs.harvard.edu/abs/2008ApJ...677.1324T} {677, 1324}

\bibitem[\protect\citeauthoryear{{Triaud}}{{Triaud}}{2018}]{Triaud2018}
{Triaud} A. H.~M.~J.,  2018, in {Deeg} H.~J.,  {Belmonte} J.~A.,  eds, , Handbook of Exoplanets.
p.~2, \mn@doi{10.1007/978-3-319-55333-7_2}

\bibitem[\protect\citeauthoryear{{Veras}}{{Veras}}{2016}]{Veras2016}
{Veras} D.,  2016, \mn@doi [Royal Society Open Science] {10.1098/rsos.150571}, \href {https://ui.adsabs.harvard.edu/abs/2016RSOS....350571V} {3, 150571}

\bibitem[\protect\citeauthoryear{{Villaver}, {Livio}, {Mustill}  \& {Siess}}{{Villaver} et~al.}{2014}]{Villaver2014}
{Villaver} E.,  {Livio} M.,  {Mustill} A.~J.,   {Siess} L.,  2014, \mn@doi [\apj] {10.1088/0004-637X/794/1/3}, \href {https://ui.adsabs.harvard.edu/abs/2014ApJ...794....3V} {794, 3}

\bibitem[\protect\citeauthoryear{{Vowell} et~al.,}{{Vowell} et~al.}{2025}]{Vowell2025}
{Vowell} N.,  et~al., 2025, \mn@doi [\aj] {10.3847/1538-3881/addd17}, \href {https://ui.adsabs.harvard.edu/abs/2025AJ....170...68V} {170, 68}

\bibitem[\protect\citeauthoryear{{Weiss} et~al.,}{{Weiss} et~al.}{2013}]{Weiss2013}
{Weiss} L.~M.,  et~al., 2013, \mn@doi [\apj] {10.1088/0004-637X/768/1/14}, \href {https://ui.adsabs.harvard.edu/abs/2013ApJ...768...14W} {768, 14}

\bibitem[\protect\citeauthoryear{{Yu}, {Huber}, {Bedding}  \& {Stello}}{{Yu} et~al.}{2018}]{Yu2018}
{Yu} J.,  {Huber} D.,  {Bedding} T.~R.,   {Stello} D.,  2018, \mn@doi [\mnras] {10.1093/mnrasl/sly123}, \href {https://ui.adsabs.harvard.edu/abs/2018MNRAS.480L..48Y} {480, L48}

\bibitem[\protect\citeauthoryear{{Zechmeister, M.} \& {Kürster, M.}}{{Zechmeister, M.} \& {Kürster, M.}}{2009}]{GLS}
{Zechmeister, M.} {Kürster, M.} 2009, \mn@doi [A\&A] {10.1051/0004-6361:200811296}, 496, 577

\makeatother
\end{thebibliography}

\newpage
\appendix
\section{Appendix}
\onecolumn
\begin{table}
\caption{Median values and 68\% confidence interval for TOI-6884 for both the probability solutions.}
\centering
\begin{tabular}{llcc}
\hline
Parameter & Description & \textbf{Value ($\Pr$ $\sim$77\%, high-mass)} & Value ($\Pr$ $\sim$23\%, low-mass) \\
\hline
\multicolumn{4}{l}{\textbf{Stellar Parameters:}} \\[2pt]
$M_*$ & Mass (\msun)    & $1.410^{+0.075}_{-0.069}$ & $1.212^{+0.035}_{-0.041}$ \\
$R_*$ & Radius (\rsun)    & $1.840^{+0.072}_{-0.073}$ & $1.868^{+0.060}_{-0.061}$ \\
$L_*$ & Luminosity (\lsun)    & $4.91^{+0.45}_{-0.42}$ & $4.57^{+0.39}_{-0.32}$ \\
$\rho_*$ & Density (cgs)    & $0.318^{+0.048}_{-0.037}$ & $0.261^{+0.026}_{-0.023}$ \\
$\log{g}$ & Surface gravity (cgs)    & $4.057^{+0.045}_{-0.039}$ & $3.977^{+0.028}_{-0.027}$ \\
$T_{\rm eff}$ & Effective temperature (K)    & $6330^{+180}_{-160}$ & $6180^{+140}_{-130}$ \\
$[{\rm Fe/H}]$ & Metallicity (dex)    & $0.094^{+0.073}_{-0.068}$ & $0.067^{+0.070}_{-0.061}$ \\
$Age$ & Age (Gyr)    & $2.61^{+0.78}_{-0.75}$ & $5.20^{+0.76}_{-0.55}$ \\
$EEP$ & Equal Evolutionary Phase     & $384^{+22}_{-29}$ & $448.3^{+5.2}_{-4.0}$ \\
$A_V$ & V-band extinction (mag)    & $0.211^{+0.094}_{-0.098}$ & $0.128^{+0.090}_{-0.081}$ \\
$\sigma_{SED}$ & SED photometry error scaling     & $0.55^{+0.23}_{-0.14}$ & $0.60^{+0.26}_{-0.16}$ \\
$\varpi$ & Parallax (mas)    & $1.778\pm0.022$ & $1.781\pm0.022$ \\
$d$ & Distance (pc)    & $562.5^{+7.1}_{-7.0}$ & $561.5^{+6.9}_{-6.8}$ \\

\hline
\multicolumn{4}{l}{\textbf{Planetary Parameters:}} \\[2pt]
$P$, b & Period (days)    & $4.808264^{+0.000015}_{-0.000014}$ & $4.808264^{+0.000015}_{-0.000014}$ \\
$R_P$, b & Radius (\rj)    & $0.927^{+0.051}_{-0.052}$ & $0.961^{+0.042}_{-0.043}$ \\
$M_P$, b & Mass (\mj)    & $26.32^{+0.98}_{-0.93}$ & $23.84^{+0.54}_{-0.64}$ \\
$T_C$, b & Observed Time of conjunction (\bjdtdb)    & $2459641.0987^{+0.0026}_{-0.0028}$ & $2459641.0988^{+0.0026}_{-0.0029}$ \\
$a$, b & Semi-major axis (AU)    & $0.0629^{+0.0011}_{-0.0010}$ & $0.05981^{+0.00057}_{-0.00067}$ \\
$i$, b & Inclination (Degrees)    & $82.46^{+0.45}_{-0.42}$ & $81.80^{+0.34}_{-0.32}$ \\
$e$, b & Eccentricity     & $0.067^{+0.010}_{-0.012}$ & $0.0684^{+0.0099}_{-0.011}$ \\
$\omega_*$, b & Arg of periastron (Degrees)    & $82.3^{+5.2}_{-6.9}$ & $82.7^{+4.9}_{-6.3}$ \\
$\dot{\omega}_{\rm GR,b}$ & Computed GR precession ($^\circ$/century)    & $1.824^{+0.064}_{-0.060}$ & $1.649^{+0.032}_{-0.037}$ \\
$T_{\rm eq,b}$ & Equilibrium temp (K)    & $1652\pm32$ & $1665^{+32}_{-28}$ \\
$\tau_{\rm circ,b}$ & Tidal circ timescale (Gyr)    & $130^{+47}_{-32}$ & $88^{+23}_{-17}$ \\
$K$, b & RV semi-amplitude (m/s)    & $2476^{+30}_{-34}$ & $2476^{+30}_{-33}$ \\
$R_P/R_*$, b & Radius of planet in stellar radii     & $0.0517\pm0.0011$ & $0.05283\pm0.00099$ \\
$a/R_*$, b & Semi-major axis in stellar radii     & $7.34^{+0.35}_{-0.30}$ & $6.87^{+0.22}_{-0.20}$ \\
$\delta$, b & $\left(R_P/R_*\right)^2$     & $0.00268^{+0.00012}_{-0.00011}$ & $0.00279^{+0.00011}_{-0.00010}$ \\
$\delta_{\rm g',b}$ & Transit depth in g' (frac)    & $0.002066^{+0.000086}_{-0.000092}$ & $0.002011^{+0.000098}_{-0.00010}$ \\
$\delta_{\rm i',b}$ & Transit depth in i' (frac)    & $0.002403^{+0.000077}_{-0.000074}$ & $0.002439^{+0.000077}_{-0.000076}$ \\
$\delta_{\rm r',b}$ & Transit depth in r' (frac)    & $0.002235\pm0.000076$ & $0.002224^{+0.000086}_{-0.000085}$ \\
$\delta_{\rm z',b}$ & Transit depth in z' (frac)    & $0.002432^{+0.000087}_{-0.000083}$ & $0.002474^{+0.000090}_{-0.000087}$ \\
$\delta_{\rm TESS,b}$ & Transit depth in TESS (frac)    & $0.002369^{+0.000093}_{-0.000089}$ & $0.00239\pm0.00010$ \\
$\tau$, b & In/egress transit duration (days)    & $0.0245^{+0.0032}_{-0.0030}$ & $0.0291^{+0.0028}_{-0.0027}$ \\
$T_{14,b}$ & Total transit duration (days)    & $0.1068^{+0.0022}_{-0.0021}$ & $0.1097\pm0.0021$ \\
$b$, b & Transit impact parameter     & $0.902^{+0.010}_{-0.013}$ & $0.9150^{+0.0063}_{-0.0075}$ \\
$\rho_P$, b & Density (cgs)    & $41.0^{+8.2}_{-6.3}$ & $33.2^{+4.9}_{-3.9}$ \\
$logg_P$, b & Surface gravity (cgs)    & $4.880^{+0.055}_{-0.050}$ & $4.805^{+0.040}_{-0.037}$ \\
$\fave$, b & Incident Flux (\fluxcgs)    & $1.68^{+0.14}_{-0.13}$ & $1.74^{+0.14}_{-0.11}$ \\
$e\cos{\omega_*,b}$ &     & $0.0089^{+0.0064}_{-0.0059}$ & $0.0086^{+0.0063}_{-0.0058}$ \\
$e\sin{\omega_*,b}$ &     & $0.066^{+0.010}_{-0.013}$ & $0.068^{+0.010}_{-0.012}$ \\
$M_P\sin i$, b & Minimum mass (\mj)    & $26.09^{+0.99}_{-0.93}$ & $23.59^{+0.54}_{-0.64}$ \\
$M_P/M_*$, b & Mass ratio     & $0.01782^{+0.00039}_{-0.00041}$ & $0.01880^{+0.00033}_{-0.00031}$ \\
\hline
\multicolumn{4}{l}{\textbf{Wavelength Parameters:}} \\[2pt]
$u_{1,g'}$ & Linear limb-darkening coeff     & $0.440^{+0.042}_{-0.041}$ & $0.458^{+0.041}_{-0.040}$ \\
$u_{1,i'}$ & Linear limb-darkening coeff     & $0.222^{+0.027}_{-0.026}$ & $0.236\pm0.025$ \\
$u_{1,r'}$ & Linear limb-darkening coeff     & $0.337^{+0.040}_{-0.039}$ & $0.355\pm0.039$ \\
$u_{1,z'}$ & Linear limb-darkening coeff     & $0.200^{+0.039}_{-0.038}$ & $0.214^{+0.038}_{-0.037}$ \\
$u_{1,TESS}$ & Linear limb-darkening coeff     & $0.245\pm0.052$ & $0.262\pm0.050$ \\
$u_{2,g'}$ & Quadratic limb-darkening coeff     & $0.254\pm0.040$ & $0.234^{+0.038}_{-0.039}$ \\
$u_{2,i'}$ & Quadratic limb-darkening coeff     & $0.308\pm0.021$ & $0.302\pm0.021$ \\
$u_{2,r'}$ & Quadratic limb-darkening coeff     & $0.345\pm0.035$ & $0.339\pm0.035$ \\
$u_{2,z'}$ & Quadratic limb-darkening coeff     & $0.323\pm0.035$ & $0.319\pm0.035$ \\
$u_{2,TESS}$ & Quadratic limb-darkening coeff     & $0.320\pm0.049$ & $0.317^{+0.048}_{-0.049}$ \\
\hline
\end{tabular}
\label{tab:final_param}
\end{table}

\newpage
\onecolumn
\begin{table}
\caption{Median values and 68\% confidence interval for TOI-6884 for both the probability solutions.}
\centering
{\renewcommand{\arraystretch}{1.3}
\begin{tabular}{llcc}
\hline\hline
Parameter & Description & Value ($\Pr$ $\sim$77\%, high-mass) & Value ($\Pr$ $\sim$23\%, low-mass) \\
\hline
\multicolumn{4}{l}{\textbf{Instrument Parameters:}} \\[2pt]
$\gamma_{\rm rel,PARAS}$ & Relative RV Offset (m/s)    & $-2350\pm120$ & $-2350\pm120$ \\
$\gamma_{\rm rel,TRES}$ & Relative RV Offset (m/s)    & $-115\pm23$ & $-114\pm22$ \\
$\sigma_J$, PARAS & RV Jitter (m/s)    & $520^{+110}_{-82}$ & $522^{+110}_{-83}$ \\
$\sigma_J$, TRES & RV Jitter (m/s)    & $56^{+34}_{-22}$ & $55^{+32}_{-21}$ \\
$\sigma_J^2$, PARAS & RV Jitter Variance     & $271000^{+120000}_{-79000}$ & $273000^{+130000}_{-80000}$ \\
$\sigma_J^2$, TRES & RV Jitter Variance     & $3200^{+4900}_{-2000}$ & $3100^{+4500}_{-1900}$ \\
\hline\\
\multicolumn{4}{l}{\textbf{Transit Parameters:}} \\[2pt]
$\sigma^{2}$, TESS UT 2022-03-14 ${(TESS)}$ & Added Variance     & $-0.000000068^{+0.000000026}_{-0.000000025}$ & $-0.000000068^{+0.000000026}_{-0.000000025}$ \\
$\sigma^{2}$, MuSCAT2 UT 2024-05-03 ${(g')}$ & Added Variance     & $-0.00002014^{+0.00000054}_{-0.00000051}$ & $-0.00002015^{+0.00000054}_{-0.00000052}$ \\
$\sigma^{2}$, MuSCAT2 UT 2024-05-03 ${(i')}$ & Added Variance     & $-0.00002014^{+0.00000053}_{-0.00000052}$ & $-0.00002015^{+0.00000053}_{-0.00000051}$ \\
$\sigma^{2}$, MuSCAT2 UT 2024-05-03 ${(r')}$ & Added Variance     & $0.00000099^{+0.00000039}_{-0.00000038}$ & $0.00000099^{+0.00000040}_{-0.00000038}$ \\
$\sigma^{2}$, MuSCAT2 UT 2024-05-03 ${(z')}$ & Added Variance     & $0.00000162^{+0.00000059}_{-0.00000056}$ & $0.00000163^{+0.00000058}_{-0.00000056}$ \\
$\sigma^{2}$, MuSCAT3 UT 2024-12-25 ${(g')}$ & Added Variance     & $0.00000093^{+0.00000016}_{-0.00000014}$ & $0.00000093^{+0.00000015}_{-0.00000014}$ \\
$\sigma^{2}$, MuSCAT3 UT 2024-12-25 ${(i')}$ & Added Variance     & $0.00000142^{+0.00000020}_{-0.00000018}$ & $0.00000142^{+0.00000020}_{-0.00000018}$ \\
$\sigma^{2}$, MuSCAT3 UT 2024-12-25 ${(r')}$ & Added Variance     & $0.00000193^{+0.00000019}_{-0.00000018}$ & $0.00000192^{+0.00000019}_{-0.00000018}$ \\
$\sigma^{2}$, MuSCAT3 UT 2024-12-25 ${(z')}$ & Added Variance     & $0.00000077^{+0.00000016}_{-0.00000014}$ & $0.00000076^{+0.00000016}_{-0.00000014}$ \\
$\sigma^{2}$, LCO UT 2460-35-2. ${(i')}$ & Added Variance     & $-0.00000112^{+0.00000079}_{-0.00000067}$ & $-0.00000114^{+0.00000079}_{-0.00000066}$ \\
$\sigma^{2}$, LCO UT 2460-40-5. ${(i')}$ & Added Variance     & $0.00000239^{+0.00000026}_{-0.00000024}$ & $0.00000238^{+0.00000026}_{-0.00000024}$ \\
$\sigma^{2}$, LCO UT 2460-42-4. ${(i')}$ & Added Variance     & $0.00000258^{+0.00000032}_{-0.00000030}$ & $0.00000258^{+0.00000032}_{-0.00000030}$ \\
$\sigma^{2}$, LCO UT 2460-42-9. ${(i')}$ & Added Variance     & $0.00000115^{+0.00000031}_{-0.00000027}$ & $0.00000117^{+0.00000031}_{-0.00000028}$ \\
$F_0$, TESS UT 2022-03-14 ${(TESS)}$ & Baseline flux     & $1.00017\pm0.000019$ & $1.000018^{+0.000018}_{-0.000019}$ \\
$F_0$, MuSCAT2 UT 2024-05-03 ${(g')}$ & Baseline flux     & $1.000041^{+0.000089}_{-0.000088}$ & $1.000044^{+0.000090}_{-0.000089}$ \\
$F_0$, MuSCAT2 UT 2024-05-03 ${(i')}$ & Baseline flux     & $1.000110\pm0.000088$ & $1.000119^{+0.000089}_{-0.000088}$ \\
$F_0$, MuSCAT2 UT 2024-05-03 ${(r')}$ & Baseline flux     & $1.000146\pm0.000081$ & $1.000143\pm0.000082$ \\
$F_0$, MuSCAT2 UT 2024-05-03 ${(z')}$ & Baseline flux     & $1.000441\pm0.000096$ & $1.000448^{+0.000098}_{-0.000097}$ \\
$F_0$, MuSCAT3 UT 2024-12-25 ${(g')}$ & Baseline flux     & $0.999690\pm0.000089$ & $0.999694^{+0.000090}_{-0.000089}$ \\
$F_0$, MuSCAT3 UT 2024-12-25 ${(i')}$ & Baseline flux     & $0.999922\pm0.000086$ & $0.999936\pm0.000086$ \\
$F_0$, MuSCAT3 UT 2024-12-25 ${(r')}$ & Baseline flux     & $1.000070\pm0.000082$ & $1.000066^{+0.000084}_{-0.000083}$ \\
$F_0$, MuSCAT3 UT 2024-12-25 ${(z')}$ & Baseline flux     & $0.999789\pm0.000089$ & $0.999799^{+0.000090}_{-0.000089}$ \\
$F_0$, LCO UT 2460-35-2. ${(i')}$ & Baseline flux     & $0.99999\pm0.00021$ & $1.00000\pm0.00021$ \\
$F_0$, LCO UT 2460-40-5. ${(i')}$ & Baseline flux     & $0.999933\pm0.000084$ & $0.999936^{+0.000085}_{-0.000086}$ \\
$F_0$, LCO UT 2460-42-4. ${(i')}$ & Baseline flux     & $1.000024^{+0.000099}_{-0.000098}$ & $1.000030\pm0.000099$ \\
$F_0$, LCO UT 2460-42-9. ${(i')}$ & Baseline flux     & $0.99963\pm0.00012$ & $0.99964\pm0.00012$ \\
\hline
\end{tabular}
}
\label{tab:final_param1}
\end{table}
\twocolumn
\begin{figure*} 
    \centering
    \begin{subfigure}[b]{1.1\columnwidth} 
        \includegraphics[width=\linewidth]{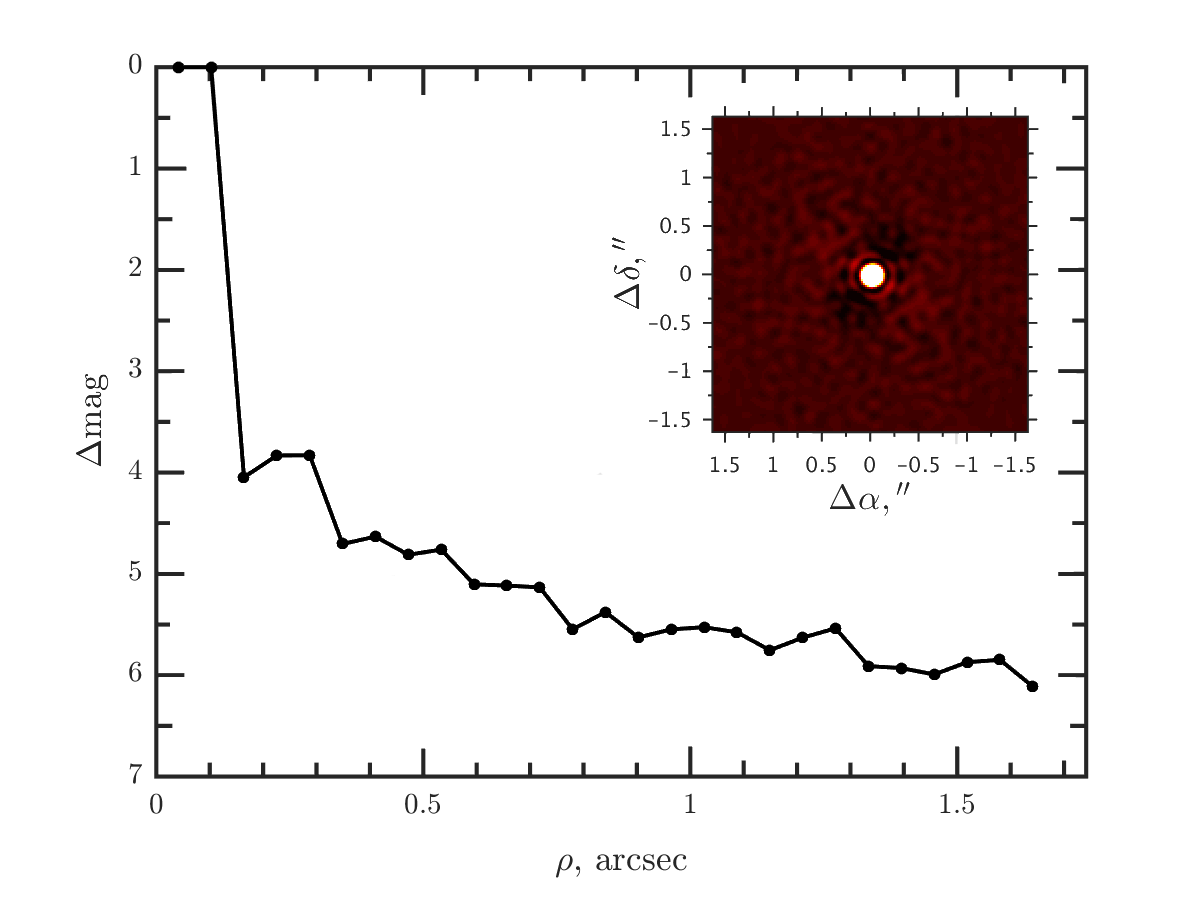}
        \subcaption{The 5$\sigma$ detection limit (contrast curve) for speckle interferometric observation of TOI-6884 with speckle polarimeter at SAI 2.5-m telescope. The inset panel shows the autocorrelation function.}
        \label{fig:toi6884_sai}
    \end{subfigure}
    \begin{subfigure}[b]{1.0\columnwidth} 
        \includegraphics[width=\linewidth]{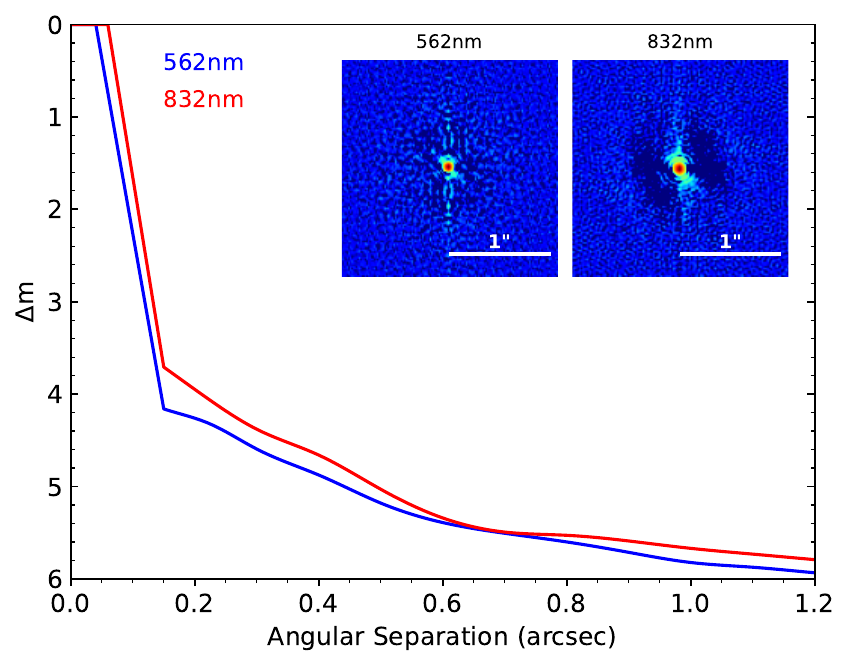}
        \subcaption{The contrast curves at 562~nm (blue curve) and 832~nm (red curve) measured from the reconstructed speckle images taken with NESSI at the WIYN 3.5~m telescope. The reconstructed speckle images are also shown as insets on the plot.}
        \label{fig:toi6884_wiyn}
    \end{subfigure}
    \caption{Speckle interferometric contrast curves for TOI-6884 obtained from two independent instruments: observations from the SAI 2.5-m telescope (top) and from the NESSI instrument on the WIYN 3.5-m telescope (bottom) (see Section ~\ref{subsec:high_imaging}).}
\end{figure*}

\begin{figure*}
    \centering
    \includegraphics[width=\linewidth]{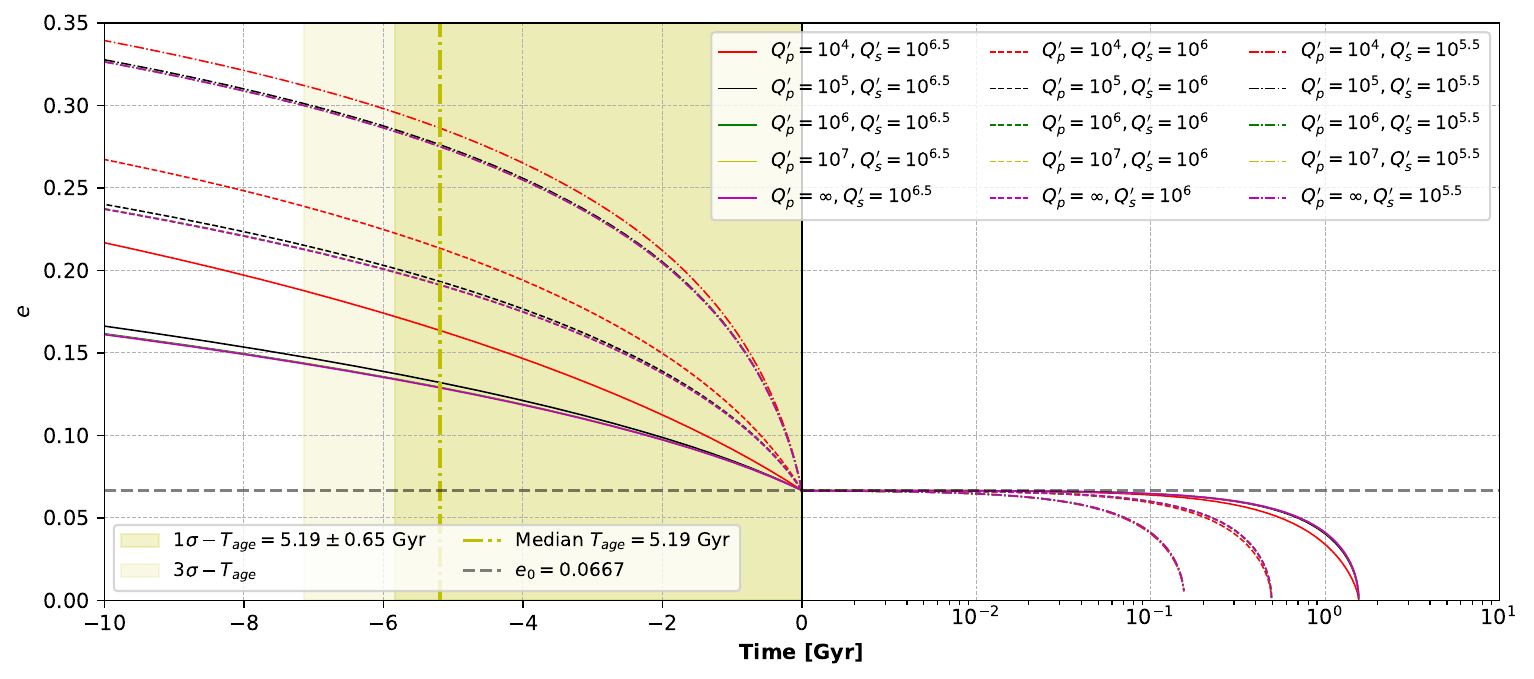}
    \caption{Backward and forward evolution of the orbital separation for different values of $Q_\star'$ and $Q_p'$, based on the formulation of \citet{Jackson+2009}, for the low-probability solution of the TOI-6884 system. The high-probability solution is shown in Fig.~\ref{fig:e-Qs-full}.}
    \label{fig:e-Qs-full_low}
\end{figure*}

\end{document}